\documentclass[11pt,a4paper]{article}
\pdfoutput=1  
\usepackage{epsfig}
\usepackage{graphicx,psfrag}
\usepackage{multirow}
\usepackage{slashed}
\usepackage{pstricks}
\usepackage{caption}
\usepackage{subcaption}
\usepackage{url}
\usepackage{braket}
\usepackage{jheppub}
\usepackage{enumerate}
\usepackage{wasysym} 
\usepackage{mathrsfs} 
\usepackage{amsfonts} 
\usepackage{amsbsy} 
\usepackage{amscd}
\usepackage{color}
\usepackage{changepage}
\usepackage{floatrow}

\makeatletter

\def\eeq{\end{equation}}
\def\beq{\begin{equation}}

\newcommand{\Rmnum}[1]{\expandafter\@slowromancap\romannumeral #1@}

\newcommand{\bea} {\begin{eqnarray}}
\newcommand{\eea} {\end{eqnarray}}

\newcommand{\gsim}{\raisebox{-0.13cm}{~\shortstack{$>$ \\[-0.07cm]
      $\sim$}}~}

\makeatother

\title{Collider Signatures of Type-X 2HDM + scalar singlet dark matter at HL-LHC}

\author[a]{Atri Dey,}
  \affiliation[a]{Regional Centre for Accelerator-based Particle Physics,
Harish-Chandra Research Institute, HBNI,
Chhatnag Road, Jhunsi, Allahabad - 211 019, India}

\author[b]{Jayita Lahiri}
   \affiliation[b]{Department of Physics, Indian Institute of Technology Guwahati,
North Guwahati, Assam - 781039, India}

\emailAdd{atridey@hri.res.in}
\emailAdd{jayitalahiri@rnd.iitg.ac.in}

\abstract{
As the 125 GeV Higgs becomes disfavored as a portal to the dark sector, one is motivated to look beyond the SM-Higgs sector, into extended scalar-mediated portal mechanisms. In this work we consider one interesting possibility of such extended scalar sector, namely Type X two Higgs doublet with a scalar singlet dark matter. This model with the advantage of light pseudoscalars also provides the explanation for the observed anomalous magnetic moment of muon. The dark matter phenomenology of this model is unique from the other types of two Higgs doublet models. Therefore, it is quite intriguing to look for signatures specific to this model at the collider experiments. In this work we take up the task of finding suitable final states and regions of parameters space that can be probed at the high-luminosity runs of LHC. We go beyond our rectangular cut-based approach and use Artificial Neural Network which shows remarkable improvement in terms of signal significance. 

}

\preprint{HRI-RECAPP-2021-014\\$\textrm{}$}

\DeclareUnicodeCharacter{2212}{-}
\begin{document}

\maketitle

\newpage

\section{Introduction}
\label{sec1}

There have been strong indirect evidence of the existence of dark matter(DM) in the universe from various astrophysical and cosmological observations. The most important question arises, what constitutes DM? A simplest extension of the SM particle spectrum by a scalar gauge-singlet makes a worthy candidate, while the 125 GeV Higgs acts as the portal to this DM~\cite{Djouadi:2012zc,Han:2016gyy}. However, such a simple extension, namely the Higgs-portal scenario, is largely disfavored, when the direct dectection bound and observed relic density are both taken into account~\cite{Lopez-Honorez:2012tov,Greljo:2013wja,Fedderke:2014wda}. The reason behind this is the fact that, in simple Higgs-portal scenarios, a single coupling takes part in the scattering as well as annhilation process. 

A solution to this to extend the scalar sector further with additional scalar doublets~\cite{Gunion:1989we,Branco:2011iw} or triplets~\cite{Magg:1980ut,Schechter:1980gr,Lazarides:1980nt,Mohapatra:1980yp,Cheng:1980qt,Bilenky:1980cx,Kobzarev:1980nk,Gunion:1989ci,Mukhopadhyaya:1990up,Mohapatra:1991ng,Ma:1998dx,Chaudhuri:2013xoa,Chaudhuri:2016rwo}. There have been studies in the past in the context of different types of Two Higgs Doublet Models (2HDM's)~\cite{Boucenna:2011hy,Berlin:2015wwa,Bandyopadhyay:2017tlq,Dey:2019lyr}, where primarily the non-standard scalar doublet acts as a portal to the dark sector. Experimental data, such as observed relic density from Planck~\cite{Ade:2013zuv}, direct detection bound from XENON-1T~\cite{Aprile:2018dbl}, Indirect detection bounds from Fermi-LAT~\cite{Wang:2017dss}, put strong constraints on the model parameter space.

The present study focusses on the Type X 2HDM with an additional scalar singlet. Type X 2HDM is special from various aspects. In this model, the non-standard scalar doublet couples only to the leptons and therefore, the gluon fusion or vector-boson fusion production of the non-standard (pseuso)scalars is highly suppressed. Hence, the lighter scalar states are admissible in this model, contrary to the other types of 2HDM's. For this special feature, Type X 2HDM garners a lot of attention from the observed muon anomaly($g_{\mu}-2$)~\cite{Bennett:2006fi,Abi:2021gix,Albahri:2021ixb}. It predicts the correct $g_{\mu}-2$ in certain regions of the model parameter space, which has interesting phenomenological consequence at the LHC~\cite{Cao:2009as,Arhrib:2011wc,Wang:2014sda,Crivellin:2015hha,Abe:2015oca,Liu:2015oaa,Chun:2017yob,Chun:2018vsn,Chakrabarty:2018qtt,Wang:2018hnw,Iguro:2019sly,Chun:2019oix,Bandyopadhyay:2019xfb,Frank:2020smf,Chun:2020uzw,Ghosh:2020tfq,Jueid:2021avn,Ghosh:2021jeg}.

On the other hand, the DM phenomenology in case of Type X 2HDM + scalar singlet is significantly different from the scenario with Type I and Type II models~\cite{Dey:2019lyr}. The preferential coupling of non-standard (pseudo)scalars to leptons, helps us evade direct detection bounds easily by making DM coupling with SM doublet small, while the correct relic density can also be achieved by tuning the couplings of DM with the other doublet. This feature sets Type X 2HDM apart from the other 2HDM's in terms of DM phenomenology. Furthermore, it is possible to have substantial invisible branching fraction for heavy scalars into DM pair, which makes it a viable scenario to look for DM at the collider.

This presents us with a great opportunity to study the collider aspects of a model, which simultaneously satisfies observed $g_{\mu}-2$ as well as provides a valid DM candidate. We perform an exhaustive scan of the parameter space with right- and wrong-sign of lepton Yukawa couplings as well as decoupling and non-decoupling extended scalar sector. We first identify the regions of parameter space which are allowed by all the relevant constraints, and choose a few suitable benchmarks from the allowed region. The benchmark selection is strongly motivated by the prospect of detectibility of this scenario at the high-luminosity LHC(HL-LHC). We also consider suitable production process and final states from this perpective. We first perform a rectangular cut-based analysis on such channels and further employ Artificial Neural Network(ANN) to achieve improved results.

The plan of this paper is as follows. In Section~\ref{sec2}, we describe the model and important aspects of it. In Section~\ref{sec3}, we discuss the theoretical, experimental and dark matter constraints and identify the allowed parameter space. Next, we choose a few suitable benchmarks for our collider analysis in Section~\ref{sec4}. In Section~\ref{sec5}, we present the cut-based analyis. The improvement over the cut-based results using machine-learning techniques, namely ANN has been explored in Section~\ref{sec6}. We summarize and conclude our analysis in Section~\ref{sec7}.

\section{Type X 2HDM with a singlet scalar}
\label{sec2}

In this work, Type X 2HDM with the inclusion of a singlet scalar $\chi$ is considered. $\chi$ can act as a possible DM candidate. The interactions of DM $\chi$ with SM particles, as well as DM self-interaction are given in the following Lagrangian.

\begin{equation}
{\cal L} = {\cal L}_{2HDM} + {\cal L}_{DM} + {\cal L}_{Int} 
\end{equation}
\begin{equation}
{\cal L}_{DM} + {\cal L}_{Int} = \frac{1}{2} \partial^{\mu} \chi  \partial_{\mu} \chi - \frac{1}{2} m_{\chi}^2 \chi^2 - \frac{\lambda_S}{4!} \chi^4 - \lambda_{1s} \chi^2 \Phi_1^{\dagger} \Phi_1 - \lambda_{2s} \chi^2 \Phi_2^{\dagger} \Phi_2
\label{lagdm}
\end{equation}

Here $\Phi_1$ and $\Phi_2$ are Higgs doublets with $Y=1$ and $\chi$ is the singlet scalar of mass $m_{\chi}$. Furthermore, 
a ${\cal Z}_2$ symmetry is postulated, under which $\chi$ is odd and
$\Phi_{1,2}$ are even. $\chi$ does not have any vacuum expectation value(vev).

We start with a
brief discussion on the different components of ${\cal L}_{2HDM}$ one by one.

\bigskip
\noindent
{\bf The scalar potential:}

\medskip
\noindent
The most general SU(2)$_L \times$ U(1)$_Y$ invariant 2HDM scalar potential is
\begin{eqnarray}
\label{lambdapotential}
\mathcal{V} &=& m_{11}^2\Phi_1^\dagger\Phi_1+m_{22}^2\Phi_2^\dagger\Phi_2
-[m_{12}^2\Phi_1^\dagger\Phi_2+{\rm h.c.}]
 +\frac{1}{2}\lambda_1(\Phi_1^\dagger\Phi_1)^2
+\frac{1}{2}\lambda_2(\Phi_2^\dagger\Phi_2)^2
+\lambda_3(\Phi_1^\dagger\Phi_1)(\Phi_2^\dagger\Phi_2) \nonumber \\
&+&\lambda_4(\Phi_1^\dagger\Phi_2)(\Phi_2^\dagger\Phi_1)
+\left\{\frac{1}{2}\lambda_5(\Phi_1^\dagger\Phi_2)^2
+\big[\lambda_6(\Phi_1^\dagger\Phi_1)
+\lambda_7(\Phi_2^\dagger\Phi_2)\big]
\Phi_1^\dagger\Phi_2+{\rm h.c.}\right\}\,. 
\end{eqnarray}

\noindent
In general, $m_{12}^2$, $\lambda_5, \lambda_6$ and $\lambda_7$ can be complex. We do not consider CP-violation
in the scalar sector in our work and hence all the parameters of the scalar potential are taken to be real.  

In order to avoid tree-level Flavor-changing-neutral-Current (FCNC)
a softly-broken ${\cal Z}^\prime_2$ symmetry is imposed on the Higgs potential implying 
$\lambda_6=\lambda_7 = 0$. After
electroweak symmetry breaking, both $\Phi_1$ and $\Phi_2$ acquire non-zero vev as follows.

\begin{equation}
\langle \Phi_1 \rangle={1\over\sqrt{2}} \left(
\begin{array}{c} 0\\ v_1\end{array}\right), \qquad \langle
\Phi_2\rangle=
{1\over\sqrt{2}}\left(\begin{array}{c}0\\ v_2
\end{array}\right)\,.\label{potmin}
\end{equation}
\noindent
where $\tan \beta = \frac{v_2}{v_1}$.

After diagonalising of the CP-even neutral scalar mass-matrix one obtains the following
physical states.

\begin{eqnarray*}
H &=&(\sqrt{2}{\rm Re\,}\Phi_1^0-v_1)\cos \alpha+
(\sqrt{2}{\rm Re\,}\Phi_2^0-v_2)\sin \alpha\,, \\
h &=&-(\sqrt{2}{\rm Re\,}\Phi_1^0-v_1)\sin \alpha+
(\sqrt{2}{\rm Re\,}\Phi_2^0-v_2)\cos \alpha\,,
\label{hZ2scalareigenstates}
\end{eqnarray*}

\noindent
The mixing angle between the CP-even scalar states is $\alpha$. In this work, we will assume that the $h$ corresponds to the SM-like Higgs and $H$ is the non-standard CP-even scalar. Both of these scalars can act as a portal to the dark sector.

\medskip
\noindent
{\bf Gauge interactions:}

\medskip
\noindent
The gauge interactions play crucial role in various production mechanisms of non-standard scalars and therefore are worth mentioning. The covariant kinetic-energy terms of the Lagrangian, which give rise to the gauge interactions are as follows.

\begin{equation}
{\cal L}_{gauge} = (D_{\mu} \Phi_1)^{\dagger} D^{\mu} \Phi_1 + (D_{\mu} \Phi_2)^{\dagger} D^{\mu} \Phi_2
\end{equation}

Where $D_{\mu} = \partial_{\mu} -   \frac{i}{2} g W_{\mu}^a \tau^a -  \frac{i}{2} g' B_{\mu}$. The coupling strengths of two physical CP-even scalars are as follows~\cite{Branco:2011iw}.

\begin{eqnarray*}
g_{hVV} = g_{SM} \times \sin(\beta - \alpha)  \\ 
h_{HVV} = g_{SM} \times \cos(\beta - \alpha)
\end{eqnarray*}

$g_{SM}$ is the corresponding coupling strengths of SM-Higgs.

\medskip

\noindent
{\bf Yukawa interactions:}

\medskip
\noindent
The Yukawa sector plays the essential role in distinguishing between different variants of 2HDM. The different interactions between the up-, down-type quarks and leptons and two scalar doublets classify various types of 2HDM. We focus here on Type X (lepton-specific) 2HDM, where one doublet couples to up- and down-type quarks and the other doublet couples to leptons. This can be achieved by imposing the discrete symmetry on the 
${\cal L}_{Yukawa}$, $\Phi_1 \rightarrow -\Phi_1$ and $e_R \rightarrow -e_R$.

The Yukawa Lagrangian for Type X 2HDM can be written as follows:

\bea
\label{yuk}
- {\cal L}_{Yukawa} &=&Y_{u2}\,\overline{Q}_L \, \tilde{{ \Phi}}_2 \,u_R
+\,Y_{d2}\,
\overline{Q}_L\,{\Phi}_2 \, d_R\, + \, Y_{\ell 1}\,\overline{L}_L \, {\Phi}_1\,e_R+\, \mbox{h.c.}\, \eea where
$Q_L^T=(u_L\,,d_L)$, $L_L^T=(\nu_L\,,l_L)$, and
$\widetilde\Phi_{1,2}=i\tau_2 \Phi_{1,2}^*$. $Y_{u2}$,
$Y_{d2}$ and $Y_{\ell 1}$ are the couplings of the up-, down-type quarks and leptons with the two doublets.

Furthermore, Yukawa couplings of the 125 GeV scalar($h$) (which can be derived in terms of the parameters of Equation~\ref{yuk}) here may or may not have the same sign as in the SM case~\cite{Han:2020zqg},
\bea
&&y_h^{f_i}~\times~y^{V}_h > 0~{\rm~right-sign(RS)},~~~\nonumber\\
&&y_h^{f_i}~\times~y^{V}_h < 0~{\rm~wrong-sign(WS)}.\label{wrongsign}
\eea

\noindent
Here $y_h^{f_i}$ and $y^{V}_h$ are the 125 GeV Higgs couplings to fermions and gauge-bosons respectively, normalized by their SM values. 
In Type-X 2HDM, the wrong-sign Yukawa coupling can arise only in the lepton Yukawa sector, unless $\tan \beta < 1$ region is allowed. However, the magnitudes of the couplings should be close to their SM values~\cite{Sirunyan:2018koj,Aad:2019mbh}, a limit known as alignment-limit.


\noindent
The lepton couplings and gauge couplings of the 125 GeV scalar can be written as follows:
\begin{equation}  
y_h^{\ell}=\sin(\beta - \alpha) - \cos(\beta -\alpha)\tan \beta,~~y^{V}_h\simeq \sin(\beta-\alpha) \nonumber\\
\end{equation}

\noindent
In the alignment limit $|\sin(\beta - \alpha)| \approx 1$. The possibilities that emerge depending on the sign of $\sin(\beta -\alpha)$ and range of $\tan \beta$, have been discussed in detail in ~\cite{Ghosh:2020tfq,Dey:2021pyn}. For the sake of completeness we mention the scenarios, relevant for our discussion.

\medskip
\noindent
$\bullet$ For $\sin(\beta-\alpha) < 0$, $\cos(\beta-\alpha) > 0$, $y_h^{\ell}$ takes the form $-(1+\epsilon)$. $y_h^{\ell}~\times~y^{V}_h > 0$ and it corresponds to right-sign region.  

\medskip
\noindent
$\bullet$ On the other hand, for $\sin(\beta-\alpha) > 0$, $\cos(\beta-\alpha) > 0$, and moderate $\tan \beta$, $y_h^{\ell}$ takes the form $(1-\epsilon)$. This case also corresponds to the right-sign region. 

\medskip
\noindent
$\bullet$ When $\sin(\beta-\alpha) > 0$ and $\cos(\beta-\alpha) > 0$ and $\tan \beta \gsim 10$, $y_h^{\ell}$ becomes negative and $y_h^{\ell}~\times~y^{V}_h < 0$. This scenario gives rise to wrong-sign lepton-Yukawa coupling.

$\epsilon$ is assumed to be an extremely small positive quantity. Also, changing the sign of $\cos(\beta-\alpha$) will lead to same ranges of $\tan \beta$ for both RS and WS regions.

Along with RS and WS regions there are two added possibilities. One should note that either the lightest (called Scenario 1 henceforth) or the second lightest (from now on called Scenario 2) CP-even scalar is the 125 GeV Higgs boson. We consider both possibilities in our work. We call the 125 GeV scalar as $h$ and the other non-standard scalar as $H$. 
We mention here that in Scenario 2, $y^{V}_h \simeq \cos(\beta-\alpha)$.

The main motivation of the present study is to explore DM phenomenology of this model with light (pseudo)scalars, the existence of which makes it easier to match the observed value of $g_{\mu}-2$. It has been shown that large $\tan \beta$ region is favored from this particular requirement. With increasing precision of future experiments the allowed region of parameter space is expected to shrink and it will be imperative to probe the remaining allowed region at the collider experiments.

\section{Constraints and allowed parameter space}
\label{sec3}

\subsection{Theoretical and phenomenological constraints}

The theoretical constraints comprise of vacuum stability, perturbativity and unitarity. 
We are currently concerned with these constraints at the electroweak state.
Vacuum essentially demands positivity of 
the potential at sufficiently large field values, in order to achieve boundedness. It suffices to check the quartic terms of the potential, since they dominate at large field values. Therefore the stability criterion essentially puts the following constraints on the quartic couplings~\cite{Deshpande:1977rw,Nie:1998yn,Drozd:2014yla}.


\begin{eqnarray}
& \lambda_1, \lambda_2, \lambda_S >0, \quad\lambda_3+\lambda_4-|\lambda_5|>-\sqrt{\lambda_1 \lambda_2}, \quad
\lambda_3>-\sqrt{\lambda_1 \lambda_2} & \label{first3}\\
&  \lambda_{1s} > - \sqrt{\frac{1}{12}\lambda_{S}\lambda_1}, \quad 
\lambda_{2s} > - \sqrt{\frac{1}{12}\lambda_{S}\lambda_2}\,. &
\label{kapsimp}
\end{eqnarray}

If $\lambda_{1s}~{\rm or}~ \lambda_{2s}<0$, then one needs to also satisfy
\begin{eqnarray}
& - 2\lambda_{1s}\lambda_{2s}+\frac{1}{6}\lambda_{S}\lambda_3>-\sqrt{ 4\left(\frac{1}{12}\lambda_{S}\lambda_1 - \lambda_{1s}^2 \right)\left(\frac{1}{12}\lambda_{S}\lambda_2-\lambda_{2s}^2\right)}& \label{kapcomp1}\\ 
& - 2\lambda_{1s}\lambda_{2s}+\frac{1}{6}\lambda_{S}(\lambda_3 +\lambda_4-|\lambda_5| )>-\sqrt{ 4\left(\frac{1}{12}\lambda_{S}\lambda_1 - \lambda_{1s}^2 \right)\left(\frac{1}{12}\lambda_{S}\lambda_2-\lambda_{2s}^2\right)} \,.&
\label{kapcomp2}
\end{eqnarray}

Perturbativity of the quartic couplings at electroweak scale implies~\cite{Drozd:2014yla} 

\begin{eqnarray}
C_{H_i H_j H_k H_l} < 4\pi \\
0 < \lambda_S < 4 \pi, |\lambda_{1s}| , |\lambda_{2s}| < 4\pi
\end{eqnarray}

Furthermore, tree-level unitarity of the scattering amplitudes involving the scalars and the longitudinal components of electroweak gauge bosons ~\cite{Arhrib:2000is,Kanemura:1993hm} demands that the eigen-values of the all the relevant scattering matrices are $< 16 \pi$~\cite{Lee:1977yc,Lee:1977eg,Ginzburg:2005dt}.

Next come experimental constraints from colliders. Electroweak
precision measurements~\cite{BAAK:2014gga}, especially the oblique parameters~\cite{Peskin:1991sw} put significant constraint on Type X parameter space when considered at one-loop level, because of the addition of additional scalars. It has been shown in various works that the non-standard scalar masses should be close to each other in order to avoid the custodial SU(2) at loop level~\cite{He:2001tp,Grimus:2007if,Bhattacharyya:2015nca,Erler:2019hds}. The regions of parameter space has also been identified in earlier works~\cite{Broggio:2014mna,Ghosh:2020tfq,Ghosh:2021jeg,Dey:2021pyn}.

CMS and ATLAS data from runs I and II on the observed 125-GeV scalar have been 
limiting its signal strengths in various channels with increasing precision~\cite{ATLAS:2016neq,CMS:2017rli,CMS:2017jkd,CMS:2017pzi,CMS:2017zyp,ATLAS-CONF-2017-045,ATLAS-CONF-2017-043}. The data shows increasing agreement to the SM values of couplings and we are pushed towards the
so-called alignment limit, ie, $(\beta - \alpha) = \pi/2$.

Lastly, the direct search for non-standard scalars with no positive result so far does constrain the parameter space rather strongly. A crucial point regarding Type X 2HDM is the fact that the production cross-section of $H/A$ in gluon fusion or vector boson fusion channels are suppressed by $\frac{1}{\tan^2\beta}$, the suppression being major in the large $\tan\beta$ region of our interest. Therefore, the masses of the non-standard scalars ($H,H^{\pm},A$) can in principle be small. This property makes it significantly different from other types of 2HDMs.    
However, stringent constraint comes from $H \rightarrow \tau \tau$ search since the $H\tau\tau$ coupling increases at large $\tan\beta$. 

One should also keep in mind, as low mass pseudoscalars are favored from $g_{\mu}-2$ considerations, there can be significant branching ratio of the 125 GeV Higgs to a pair of pseudoscalars as long as $m_A < \frac{m_h}{2}$. However, collider searches have put strong constraint on this branching fraction. In order to obey that limit the coupling between $h$ and a pair of pseudoscalars, $g_{hAA}$ has be extremely strong. Parameter space consistent with this constraint have been identified in~\cite{Ghosh:2020tfq,Dey:2021pyn}.

\subsection{Muon $g-2$ constraints}

It is well-known that the observed anomalous magnetic moment of muon ($g_{\mu}-2$) can not be explained in the SM~\cite{Aoyama:2020ynm,Davier:2017zfy,Keshavarzi:2018mgv,Colangelo:2018mtw,Hoferichter:2019mqg,Davier:2019can,Keshavarzi:2019abf,Kurz:2014wya,Melnikov:2003xd,Masjuan:2017tvw,Colangelo:2017fiz,Hoferichter:2018kwz,Gerardin:2019vio,Bijnens:2019ghy,Colangelo:2019uex,Colangelo:2014qya,Blum:2019ugy,Aoyama:2012wk,Czarnecki:2002nt,Aoyama:2019ryr,Gnendiger:2013pva,Zyla:2020zbs}. There are models where additional contribution to $g_{\mu}-2$ can be achieved and therefore, are of significant importance. Type X 2HDM, is one such model which can provide correct $g_{\mu}-2$ in certain regions of parameter space. However, the current precise experimental measurements of $g_{\mu}-2$~\cite{Bennett:2006fi,Abi:2021gix,Albahri:2021ixb} constrain the model parameter space to a large extent. The details of $g_{\mu}-2$ in the context of Type X 2HDM have been performed in various studies~\cite{Cao:2009as,Arhrib:2011wc,Wang:2014sda,Crivellin:2015hha,Abe:2015oca,Liu:2015oaa,Chun:2017yob,Chun:2018vsn,Chakrabarty:2018qtt,Wang:2018hnw,Iguro:2019sly,Chun:2019oix,Bandyopadhyay:2019xfb,Frank:2020smf,Chun:2020uzw,Ghosh:2020tfq,Jueid:2021avn,Ghosh:2021jeg}. Here we would only mention that the major contribution to $g_{\mu}-2$ comes from two-loop Bar-Zee diagrams involving light-non-standard (pseudo)scalars(which are allowed in Type X 2HDM as discussed above) and $\tau$-leptons. As the $\tau$-lepton coupling to (pseudo)scalars increases with $\tan\beta$, the $g_{\mu} - 2$ contribution also increases with $\tan \beta$. On the other hand, low masses of the non-standard scalars (especially the pseudoscalar A, since it has most enhanced coupling to the leptons $\propto \tan\beta$ in Type X 2HDM) are preferred from the observed $g_{\mu}-2$. The allowed regions in $m_A-\tan\beta$ space with the most updated constraints can be found in~\cite{Ghosh:2021jeg,Dey:2021pyn}.

\subsection{Constraints from dark matter search and allowed parameter space}

For $\chi$ to be considered as a thermal dark matter candidate, it has been required
here to satisfy the following constraints:

\begin{itemize}
\item The thermal relic density of $\chi$ should not exceed the latest Planck data~\cite{Ade:2013zuv} at the 2$\sigma$ level.

\item The $\chi$-nucleon cross section  should be below 
the current upper bound from XENON1T~\cite{Aprile:2018dbl}.

\item Constraints from indirect detection experiments should be satisfied. Therefore the annihilation rate of $\chi$ has been consistent at
 the 95\% confidence level with both isotropic gamma-ray distribution data
 and the gamma ray observations from dwarf spheroidal galaxies~\cite{Ackermann:2015zua}.

\item The invisible decay of the 125-GeV scalar $h$ has been limited to
19\%~\cite{Sirunyan:2018owy}.
\end{itemize}

We will identify the impacts of all these constraints on our model parameter space one by one. In order to obtain regions allowed by all the aforementioned constraints, we perform a scan of the parameter space. The ranges of scan for the two Types of models are as follows(Table.~\ref{scan}):

\begin{table}[!hptb]
\begin{center}
\begin{footnotesize}
\begin{tabular}{| c |}
\hline
 $80 \text{GeV} < m_H < 3\text{TeV}$ \\
$80 \text{GeV} < m_H^{\pm} < 3\text{TeV}$ \\
$10 \text{GeV} < m_{\chi} < 400 \text{GeV}$ \\
$-12 < \lambda_{1s} < 12$\\
$-12 < \lambda_{2s} < 12$\\
$10 < \tan \beta < 100$\\
$0.99 < |\sin (\beta - \alpha)| < 1$ \\
\hline
\end{tabular}
\end{footnotesize}
\caption{The range of scan for relevant parameters.}
\label{scan}
\end{center}
\end{table}

\begin{figure}[!hptb]
\includegraphics[width=8.0cm,height=7.0cm]{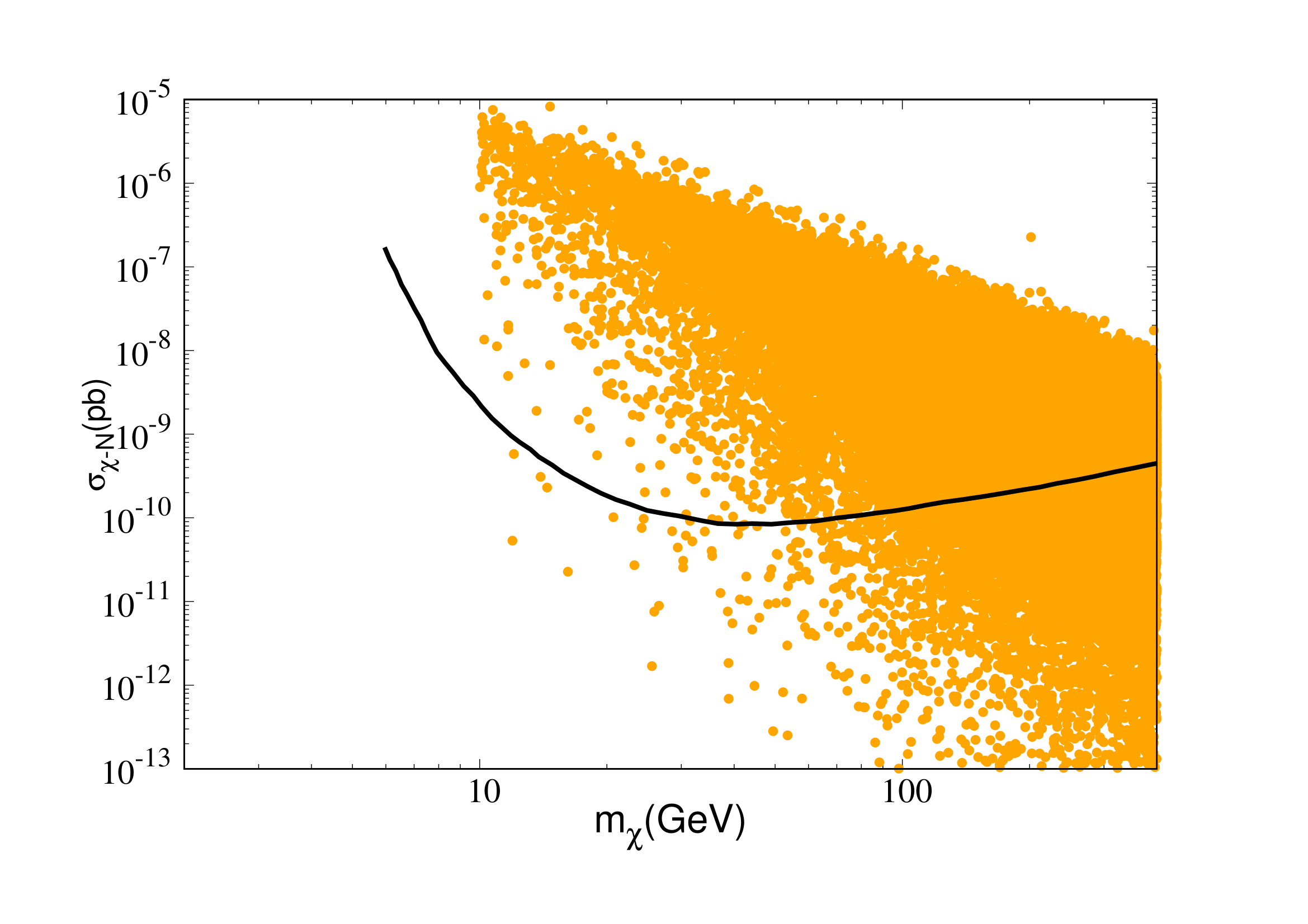}
\caption{Parameter space allowed by the requirement of relic underabundance(orange points). The black dashed line is the upper limit on the $\chi-N$ scattering cross section from XENON1T experiment.}
\label{sigma}
\end{figure}

In Figure~\ref{sigma}, we select the parameter space allowed by the upper bound of relic density(orange points) and plot the direct detection cross-section ($\sigma_{\chi-N}$) as a function of $m_{\chi}$. One can see that the cross-section decreases with increasing DM mass (see the analytical expression in Equation~\ref{sigma_eq}). We also show the upper limit from XENON-1T experiments in the same plot.

\begin{figure}[!hptb]
\includegraphics[width=7.0cm,height=6cm]{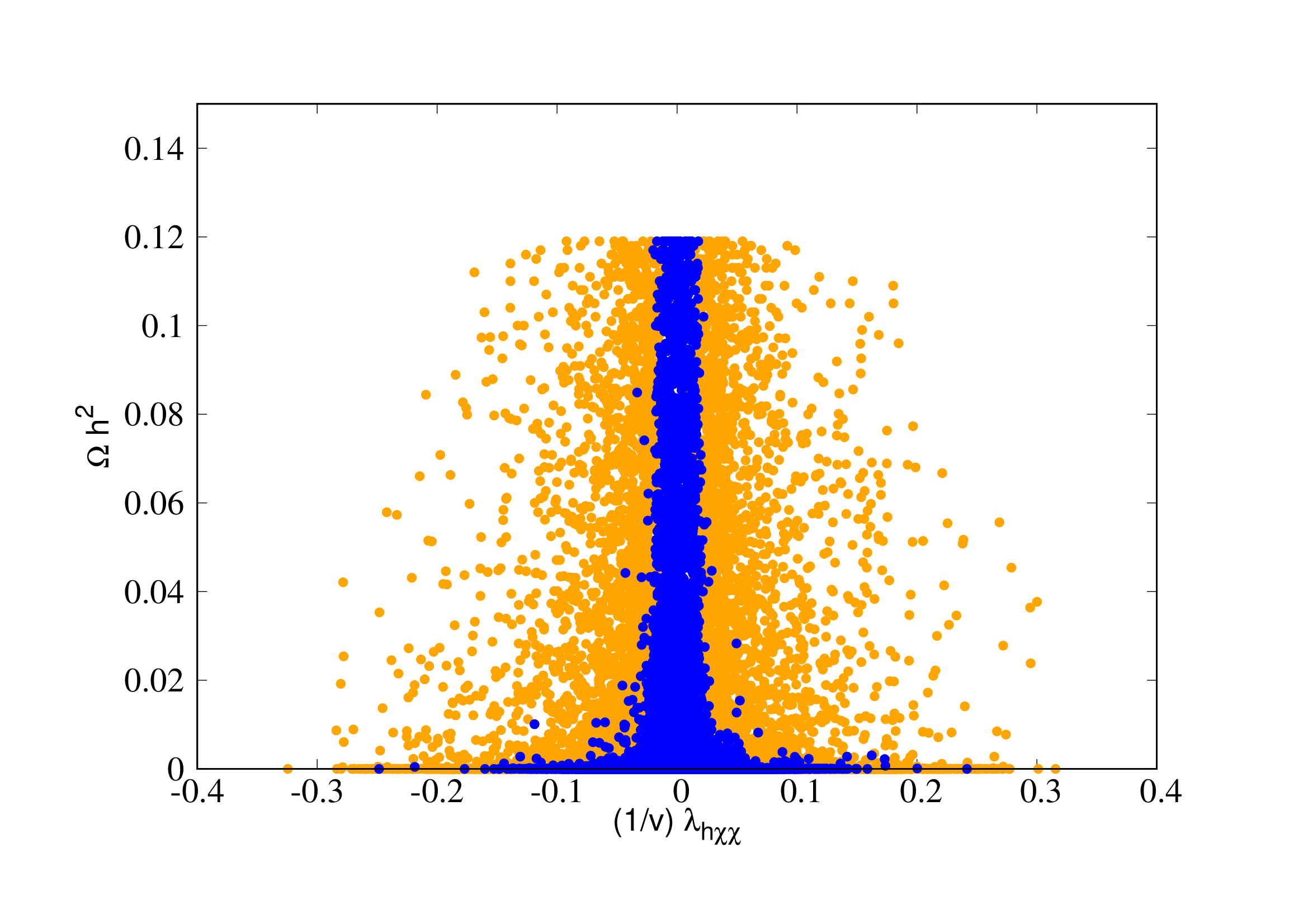}
\includegraphics[width=7.0cm,height=6cm]{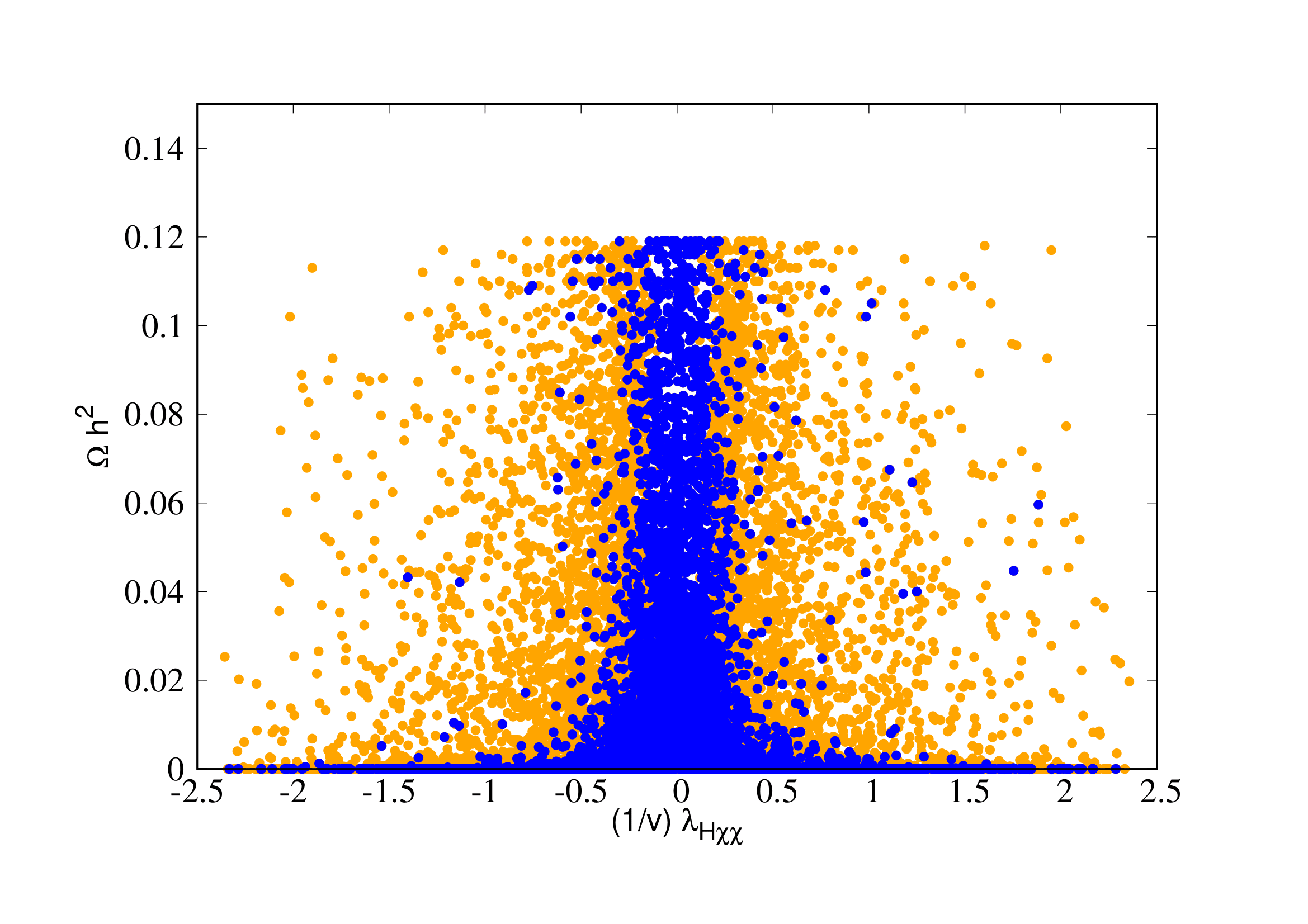}
\caption{Dependence of relic density on $\frac{1}{v}\lambda_{h\chi\chi}$ and $\frac{1}{v}\lambda_{H\chi\chi}$ 
The orange points satisfy relic-underabundance criteria and blue points satisfy direct detection constraints as well.}
\label{coupling_relic}
\end{figure}

In Figure~\ref{coupling_relic}, we show the dependence of relic density on dark matter couplings with physical scalars($h$ and $H$). The respective couplings $\lambda_{h\chi\chi}$ and $\lambda_{H\chi\chi}$ can be expressed in terms of the coupling parameters of the Lagrangian (see Equation~\ref{lagdm}) $\lambda_{1s}$ and $\lambda_{2s}$ as follows:

\begin{eqnarray}
\lambda_{h \chi \chi} =  \lambda_{1s} v \sin \alpha \cos \beta - \lambda_{2s} v\cos \alpha \sin \beta  \\
\lambda_{H \chi \chi} = -\lambda_{1s} v\cos \alpha \cos \beta - \lambda_{2s} v\sin \alpha \sin \beta 
\label{lambdahH}
\end{eqnarray}

We can see that the relic density falls with increasing coupling strength of DM with $h$ or $H$. That happens because the annihilation cross-section increases with increasing couplings. We also show the region allowed by direct detection bounds in the same plot(blue points). One can see that, the allowed couplings becomes further restricted when one applies direct detection bound. The interplay between the relic density and direct detection bound is very different in Type X 2HDM as compared to Type I and Type II 2HDM. In Type X 2DHM, the up- and down-type quarks couple to the particular doublet which couples to DM states with strength $\lambda_{1s}$ and the lepton couples to the doublet which has $\lambda_{2s}$ coupling with DM particles. Therefore, $\lambda_{1s}$ coupling takes part in $\chi$-nucleon scattering cross-section, and is constrained by its upper limits. On the other hand, $\lambda_{2s}$ remain almost unaffected by the direct detection bounds, a scenario very different from Type II 2HDM. Furthermore, the $\lambda_{1s}$ coupling which governs the annihilation of DM pair into $\tau\tau$ final states can be adjusted in order to achieve the observed relic density. It is not possible in Type I case, since in that case same coupling takes part in relic density and direct detection cross-section. It is therefore clear that in Type X 2HDM it is easier to satisfy both constraints simultaneously.

\begin{figure}[!hptb]
	\centering
	\includegraphics[width=7.5cm,height=6cm]{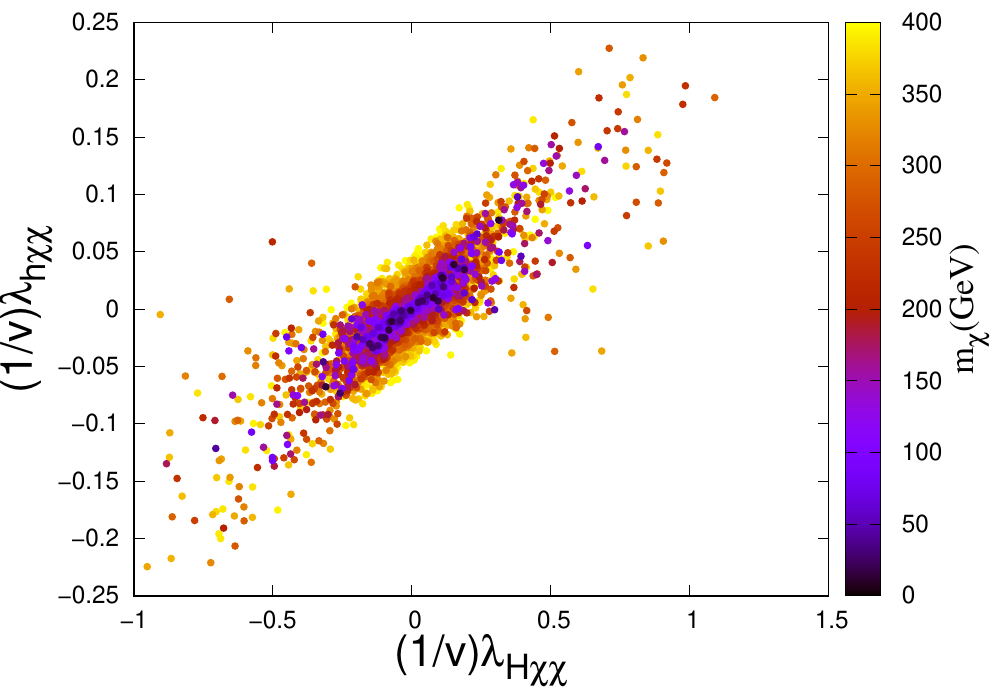}
       	\includegraphics[width=7.5cm,height=6cm]{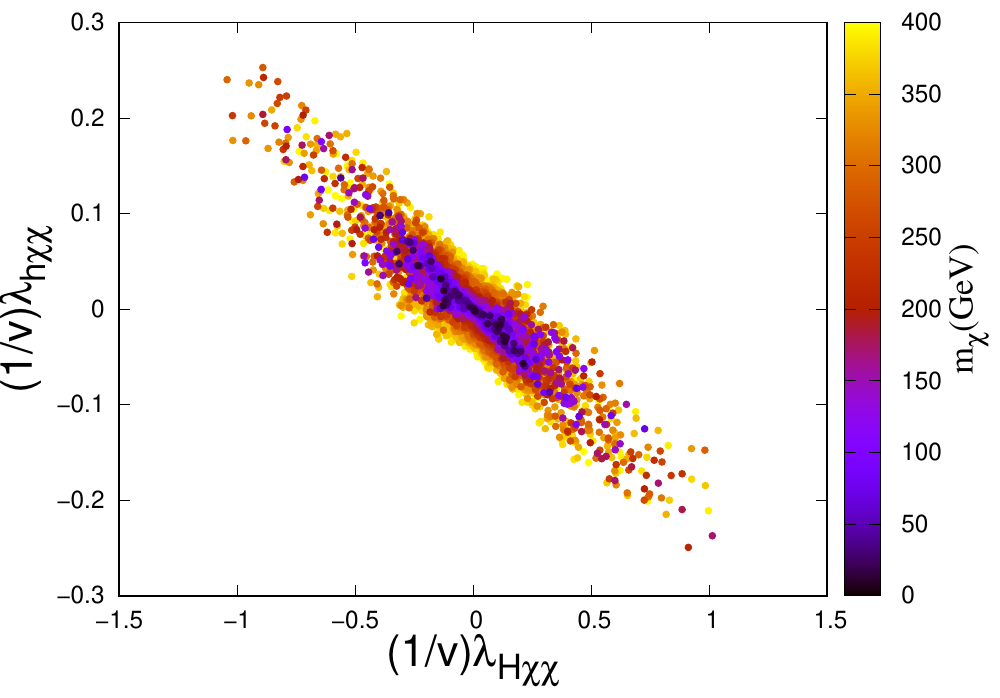}
	\caption{Parameter space allowed by direct detection bound (Xenon-1T) in the $\frac{1}{v}\lambda_{H\chi\chi}-\frac{1}{v}\lambda_{h\chi\chi}$ plane in the RS region(left) and WS region(right).}
	\label{coupling_dd}
\end{figure}

\noindent
In Figure~\ref{coupling_dd}, we show the allowed parameter space in the plane spanned by the couplings of DM with two physical scalar states, namely $\lambda_{h\chi\chi}$ and $\lambda_{H\chi\chi}$. The direct detection cross-section is given in the following form. 

\begin{equation}
\sigma_{\chi N} = \frac{m_N^2}{\pi(m_{\chi}+m_N)^2}\left(\frac{\lambda_{h\chi\chi}g_{NNh}}{m_h^2} + \frac{\lambda_{H\chi\chi}g_{NNH}}{m_H^2}\right)^2
\label{sigma_eq}
\end{equation}

\noindent
Let us first examine the couplings involved in the calculation of DM-nucleon direct detection cross-section.

\begin{eqnarray}
g_{NNh} \propto \frac{\cos\alpha}{\sin\beta} = \sin(\beta - \alpha)+\cos(\beta-\alpha)\cot\beta \\
g_{NNH} \propto \frac{\sin\alpha}{\sin\beta} = \cos(\beta - \alpha)-\sin(\beta-\alpha)\cot\beta
\end{eqnarray}

In Figure~\ref{coupling_dd}, we show the DM mass $m_{\chi}$ is shown as the color-axis. We see from Equation~\ref{sigma_eq} that the direct detection decreases with DM mass $m_{\chi}$. Also, the upper limit on direct detection cross-section becomes weaker with increasing $m_{\chi}$. Therefore, larger regions get allowed for higher DM masses, which is apparent from the color distribution in Figure~\ref{coupling_dd}.

In the right-sign case when $\sin(\beta-\alpha) > 0$ and $\sin(\beta - \alpha) - \cos(\beta - \alpha)\tan\beta > 0$, that implies $g_{NNh} > 0$ and $g_{NNH} < 0$. When $g_{NNh}$ and $g_{NNH}$ are of opposite sign, $\lambda_{h\chi\chi}$ and $\lambda_{H\chi\chi}$ have to be of same sign to ensure cancellation between the two terms in Equation~\ref{sigma_eq} and a small value of direct detection cross-section can be achieved.
On the other hand, right-sign is also possible when $\sin(\beta - \alpha) < 0$ and $\sin(\beta - \alpha) - \cos(\beta - \alpha)\tan\beta < 0$, which implies $g_{NNh} < 0$ and $g_{NNH} > 0$. Here too, the opposite sign between $g_{NNh}$ and $g_{NNh}$ will ensure same sign between $\lambda_{h\chi\chi}$ and $\lambda_{H\chi\chi}$. We can see this correlation in Figure~\ref{coupling_dd}(left).

The wrong-sign is possible only when $\sin(\beta-\alpha) > 0$ and $\sin(\beta - \alpha) - \cos(\beta - \alpha)\tan\beta < 0$. This indicates $g_{NNh} > 0$ and $g_{NNH} > 0$. Therefore the aforementioned cancellation is possible only when  $\lambda_{h\chi\chi}$ and $\lambda_{H\chi\chi}$ are of opposite sign. We can see this behavior from Figure~\ref{coupling_dd}(right).

\begin{figure}[!hptb]
	\centering
	\includegraphics[width=7.5cm,height=6cm]{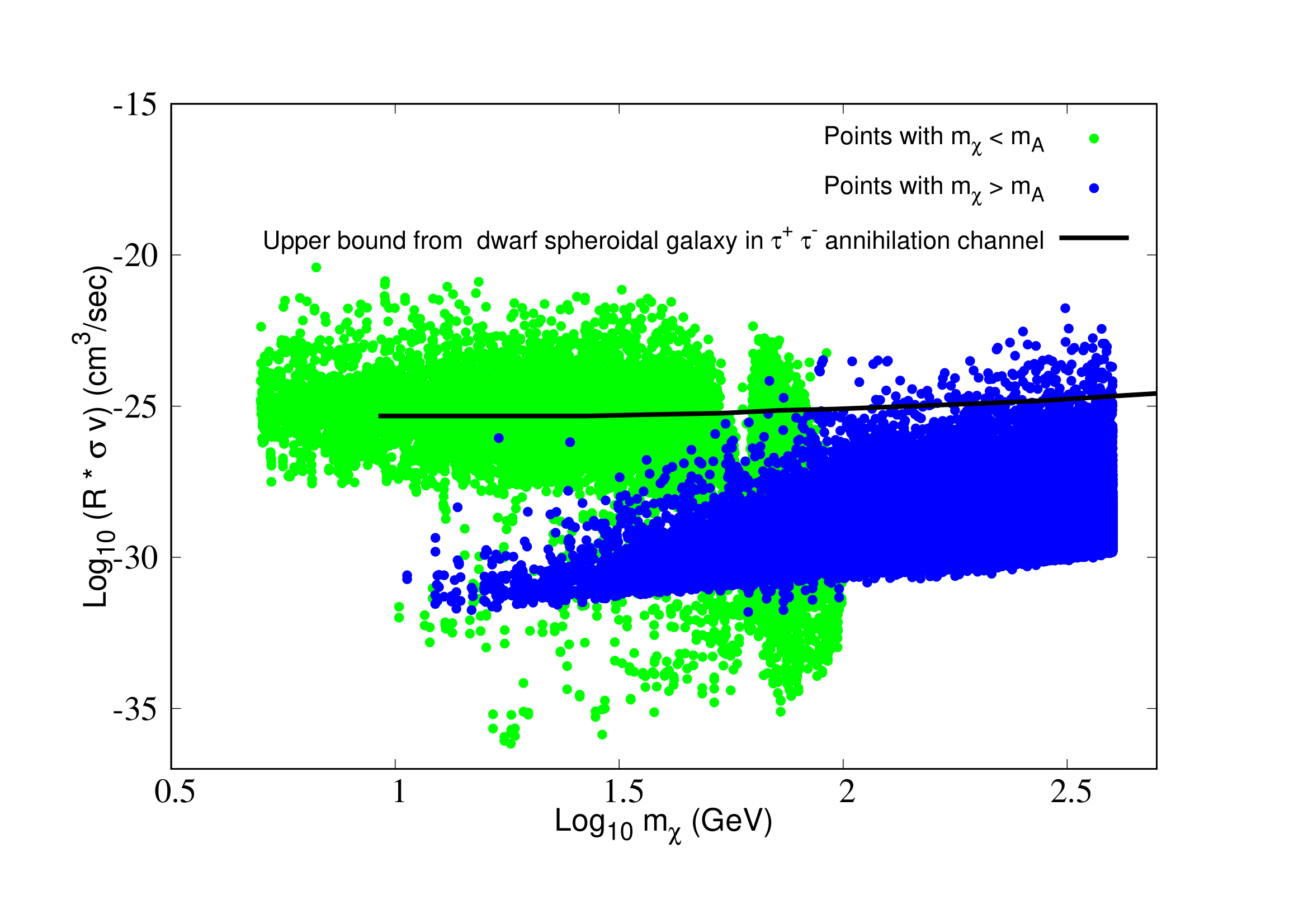}
       	\includegraphics[width=7.5cm,height=6cm]{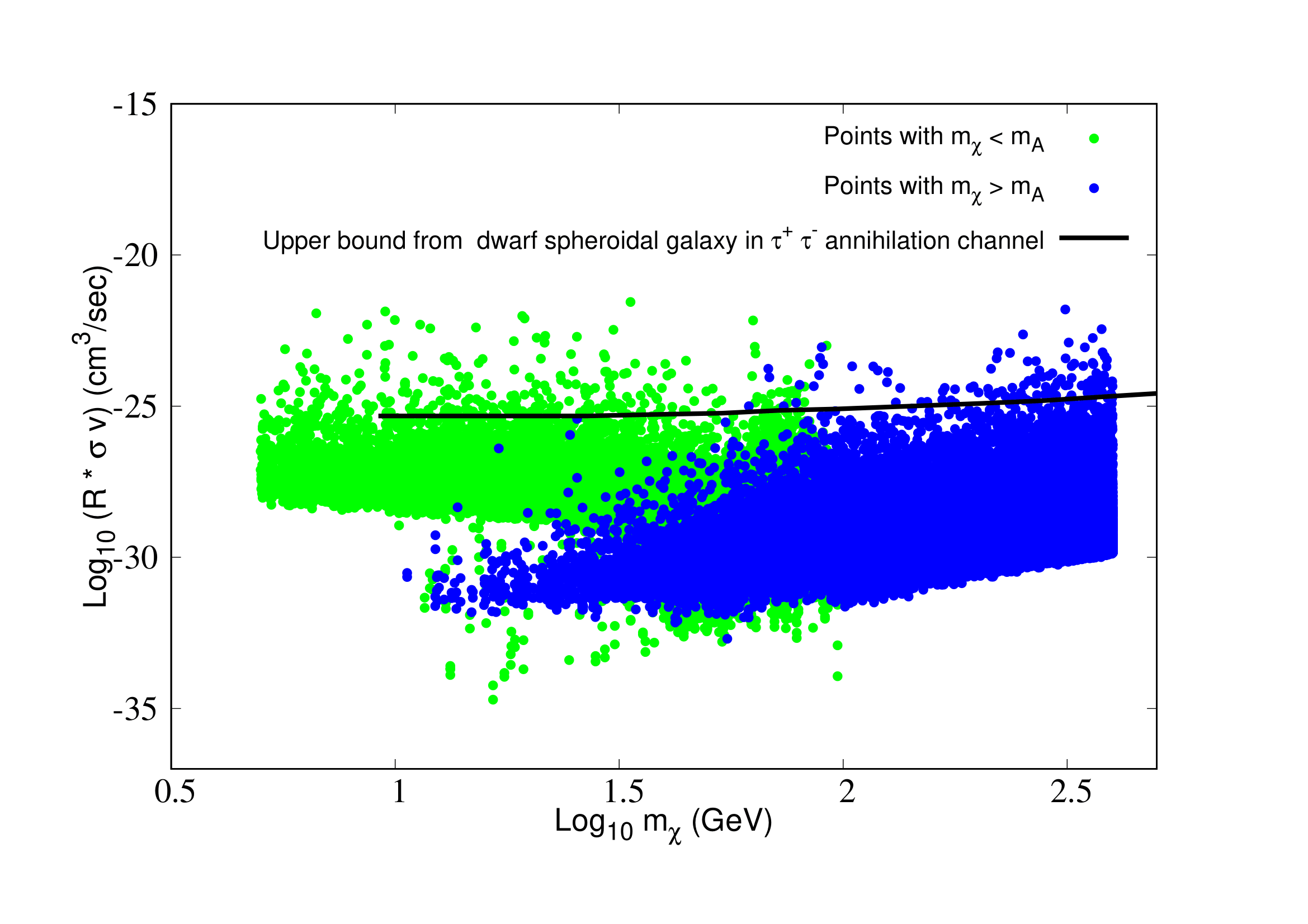}
	\caption{Parameter space allowed by indirect detection bound from Fermi-LAT collaboration.}
	\label{annihilation}
\end{figure}

One should also take into account the indirect detection constraints on the parameter space. The major annihilation channels for DM pair are $\chi\chi\rightarrow AA/H^{+}H^{-}/HH$ when these channels are kinematically feasible. When DM mass is small and these annihilations are not available, $\chi\chi\rightarrow \tau\tau$ becomes the major channel. Therefore, the major constraint comes from indirect search from DM in the $\tau\tau$ final state, especially in the low mass ($m_{\chi} < m_A$) region. In Figure~\ref{annihilation}, we show the annihilation cross-section of $\chi\chi \rightarrow \tau\tau$ as a function of $m_{\chi}$. We also show the upper limit coming from Fermi-LAT experiments in this annihilation channel, and the allowed parameter space can thus be identified.

In Figure~\ref{annihilation}, we show the distribution of annihilation cross-section for both Scenario 1(left) and 2(right). One can see that the region where $m_{\chi} < m_A$ (green points), the upper limit from $\tau\tau$ annihilation channel constrain the parameter space significantly. One should note that, in Scenario 2, the constraints are relaxed compared to Scenario 1. The reason is the following. In Scenario 2, $m_H$ and $m_H^{\pm}$ are light. Therefore, even when $m_{\chi} < m_A$, it is still possible that DM mass will annihilate into lighter $HH$ or $H^{+}H^{-}$ final states. We should mention here that while calculating the annihilation cross-section we have scaled each contribution
with their relative contribution to the DM density. In this spirit, the factor $R$ in Figure~\ref{annihilation}, is the ratio of the relic density in this model with particular set of parameters and the total observed relic density of the universe ($R=(\frac{\Omega}{\Omega_{obs}})^2$).

\begin{figure}[!hptb]
	\centering
	\includegraphics[width=8.5cm,height=7cm]{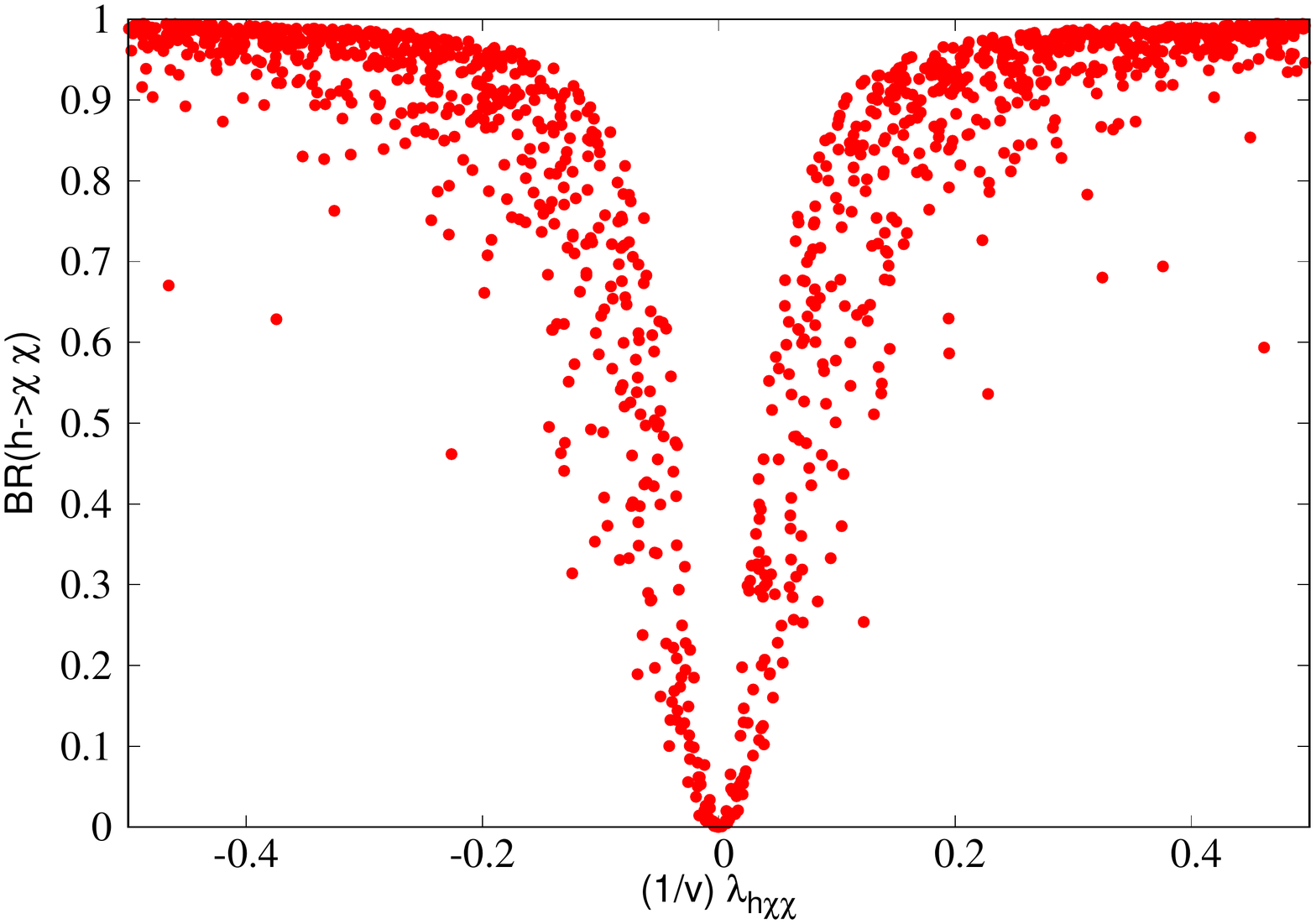}
	\caption{Br($h \rightarrow \chi\chi$) as a function of the DM-Higgs coupling $\frac{1}{v}\lambda_{h\chi\chi}$.}
	\label{brhchichi}
\end{figure}

It is very obvious that the coupling of the 125 GeV scalar to a pair of DM($\frac{1}{v}\lambda_{h\chi\chi}$) plays a crucial role in determining the branching ratio BR($h_{SM}\rightarrow\chi\chi)$. This branching ratio is highly constrained by experimental searches and the the upper limit on this is found to be 19\%~\cite{Sirunyan:2018owy}. This upper limit in turn puts a cut-off on the coupling $\frac{1}{v}\lambda_{h\chi\chi}$. In Figure~\ref{brhchichi}, we plot the dependence of BR($h \rightarrow \chi\chi$), where $h$ is the 125 GeV scalar.

\section{Selection of benchmark}
\label{sec4}

Our focus in this work is to probe the aforementioned scenario and its interesting parameter space at the future colliders. For this purpose we need to select a few representative benchmarks which satisfy all the theoretical, experimental and observational constraints. The preceding discussion helps us identify allowed regions in this context. On the other hand, another crucial requirement in this regard is to select regions of parameter space which can be probed with best possible significance at the earlier runs of HL-LHC. In Type X 2HDM, the quark couplings to the non-standard scalar states ($H,H^{\pm},A$) are suppressed. Therefore, the prominent production mechanism in a proton-proton collider will be via Drell-Yan process. Therefore, we will focus on Drell-Yan production of non-standard scalars, which is precisely determined by the gauge couplings. Secondly, the other important factor is the branching ratio(BR) of the non-standard CP-even scalar $H$ to a pair of DM particles, since the only way we can get DM in the final state at tree-level is via the decay of $H$. We will first try to concentrate on the BR($H \rightarrow \chi\chi$) in various regions of parameter space of the model. 

It is obvious that to get a large BR($H \rightarrow \chi\chi$) we must consider a large value of $\lambda_{H \chi \chi}$ which controls the coupling of DM pair to our non-standard CP-even scalar $H$. However, $m_{\chi}$, $m_H$ and $\tan\beta$ also play a crucial role here. It is discussed earlier that to satisfy observed $g_{\mu}-2$ in Type-X 2HDM, we need a moderate to large $\tan\beta$ value, which enhances the branching fraction of $H \rightarrow \tau \tau$ in this model. On the other hand, for Scenario 1, where $m_H$ can be large, BR($H \rightarrow Z A$) can become the major decay channel of $H$. 

\begin{figure}[!hptb]

\includegraphics[width=7.1cm, height=7cm]{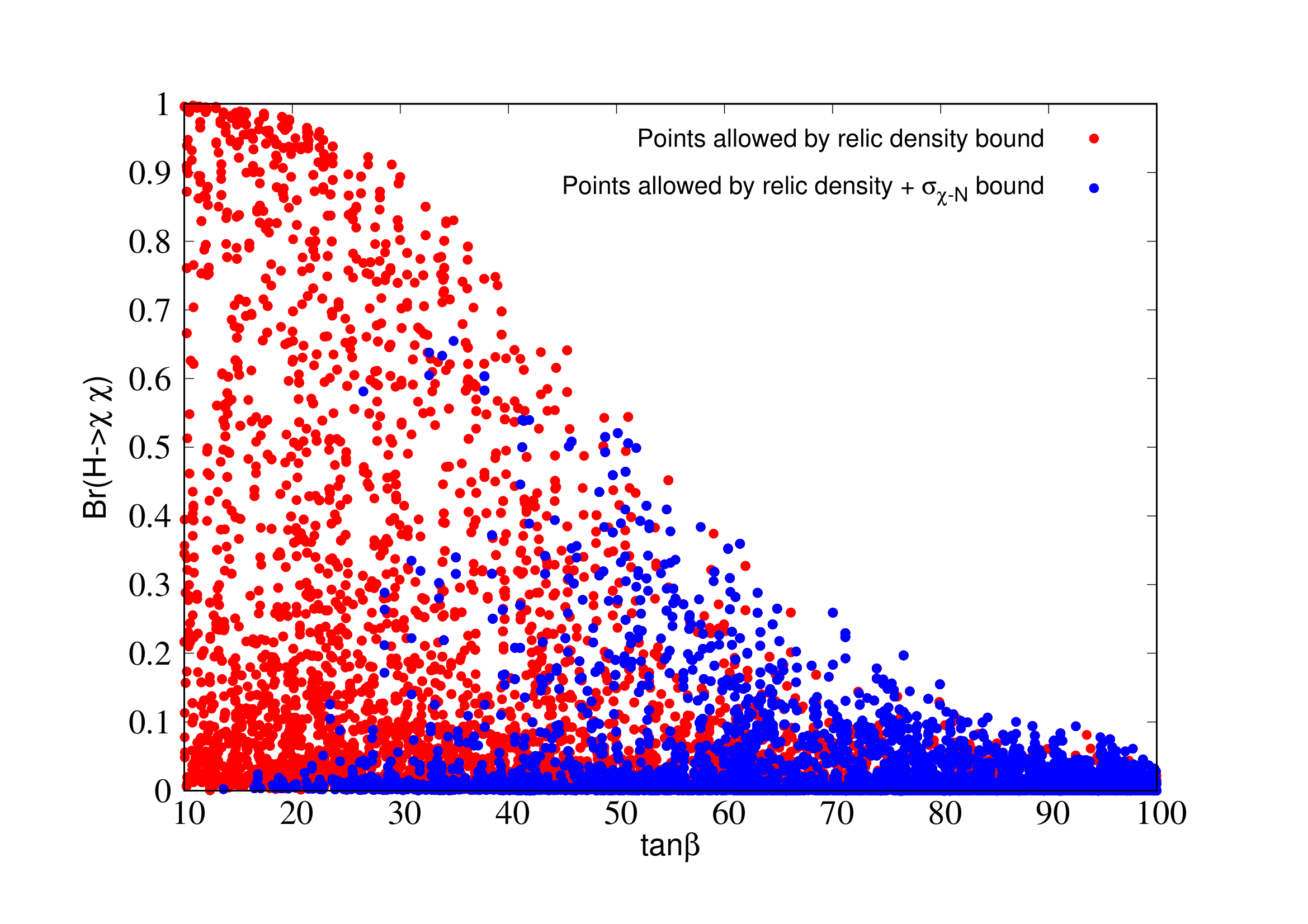}
\includegraphics[width=7.1cm, height=7cm]{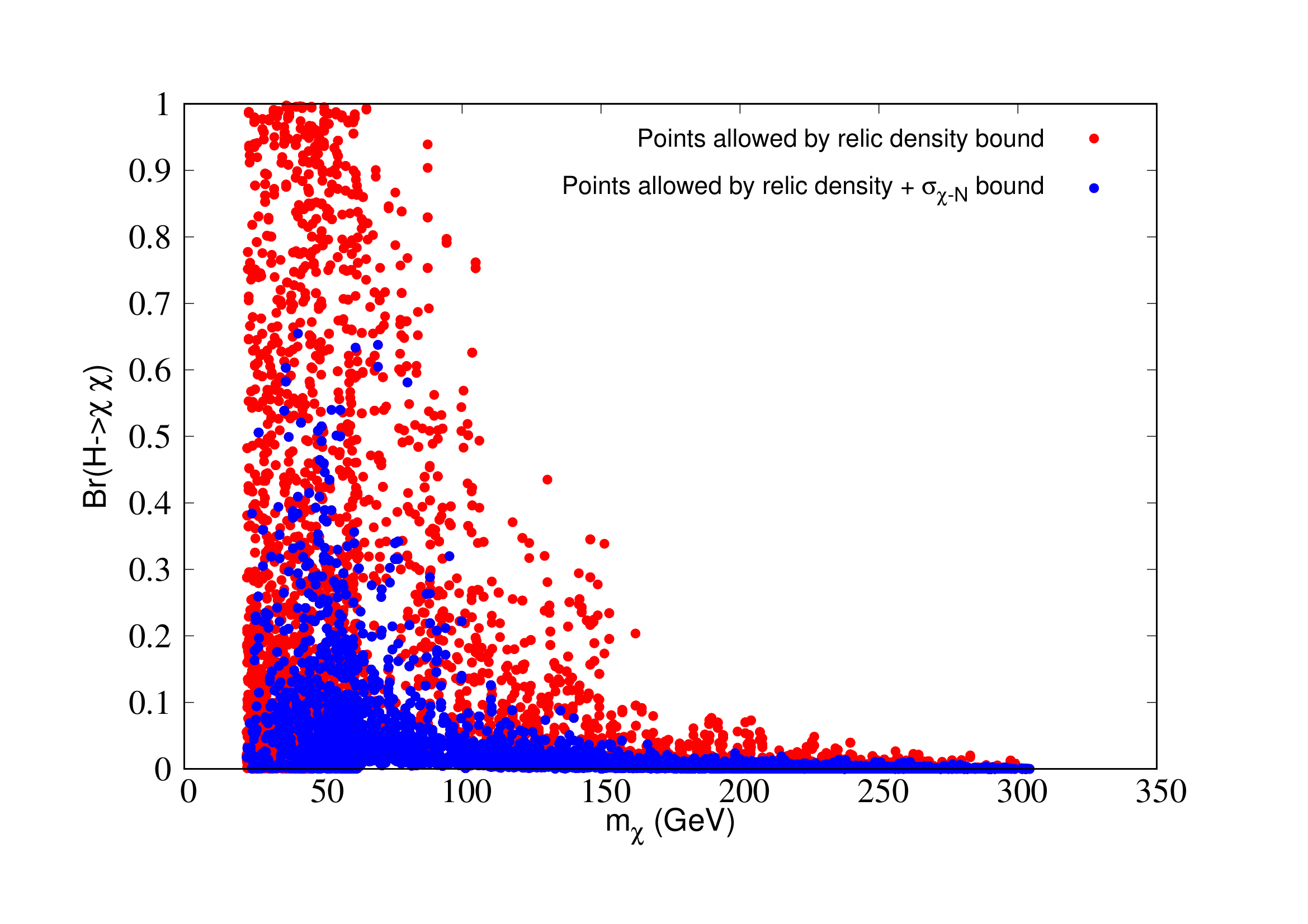}

\includegraphics[width=7.1cm, height=7cm]{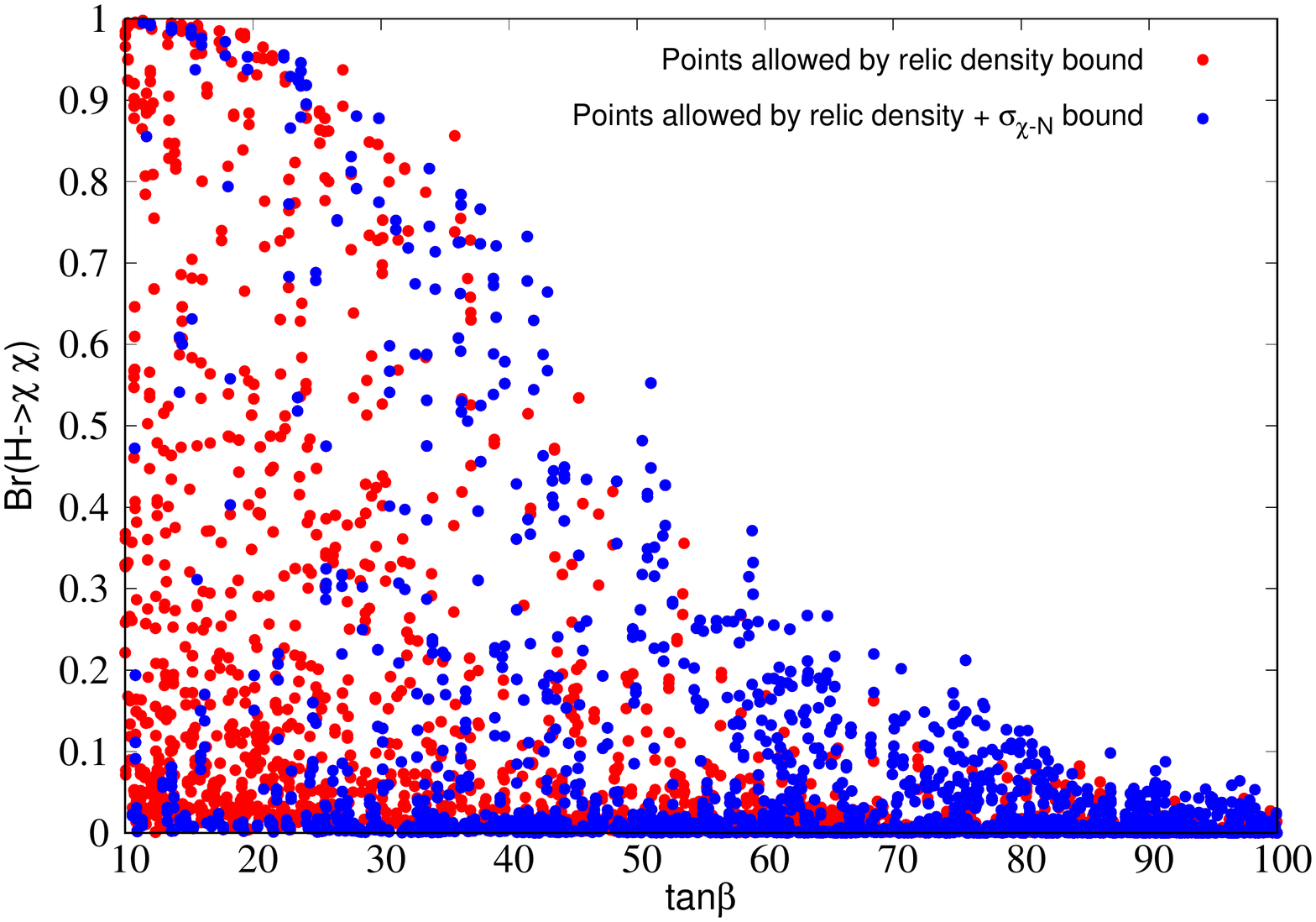}
\includegraphics[width=7.1cm, height=7cm]{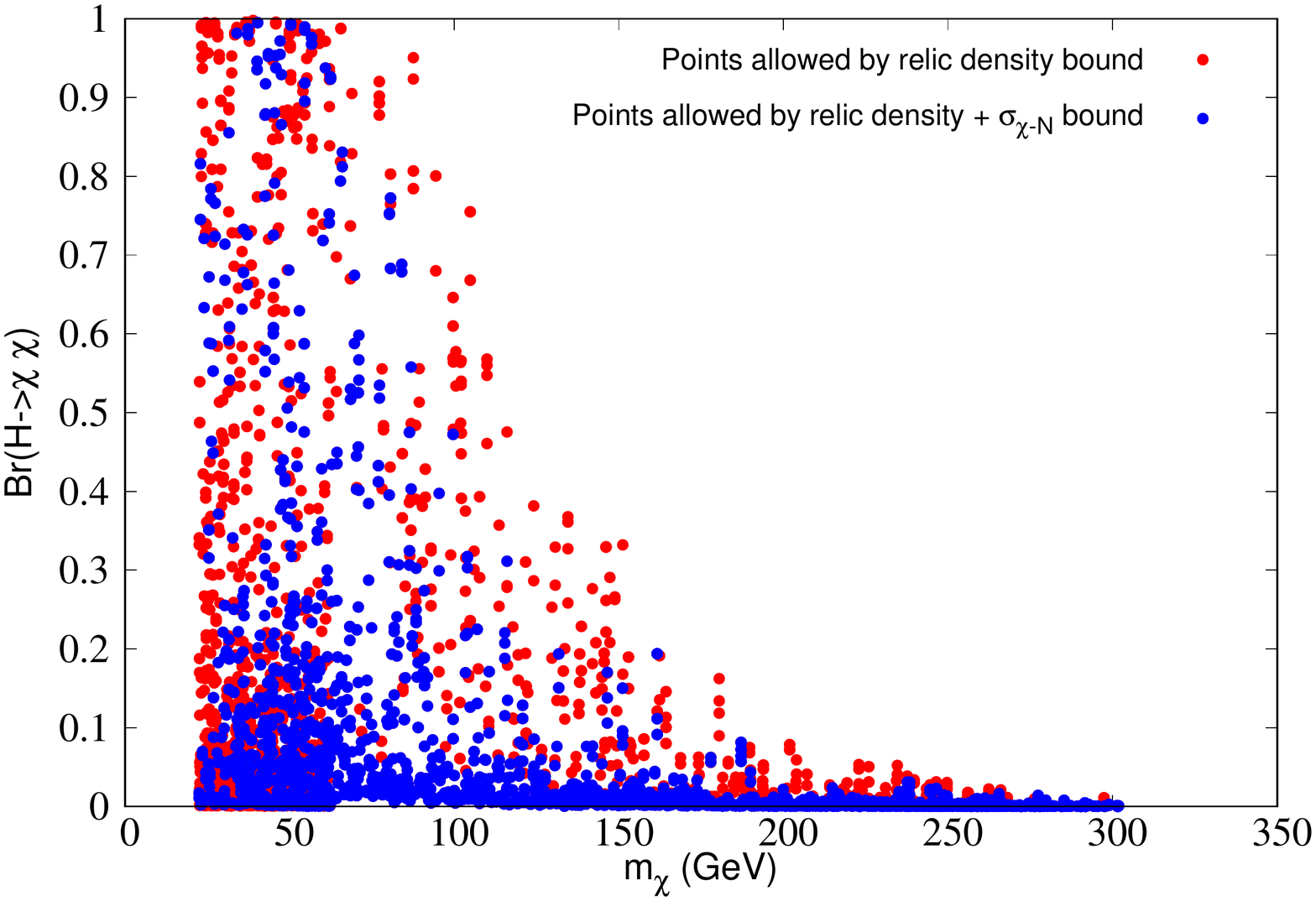}

\caption{Dependence of $BR(H \rightarrow \chi \chi)$ on $\tan \beta$(left) and $m_{\chi}$ (right) for Scenario 1. Upper two plots represent WS Yukawa case where the lower two represent RS Yukawa case.}
\label{brhxx_tb_mchi_scenario1}
\end{figure}

\begin{figure}[!hptb]

\includegraphics[width=7.1cm, height=7cm]{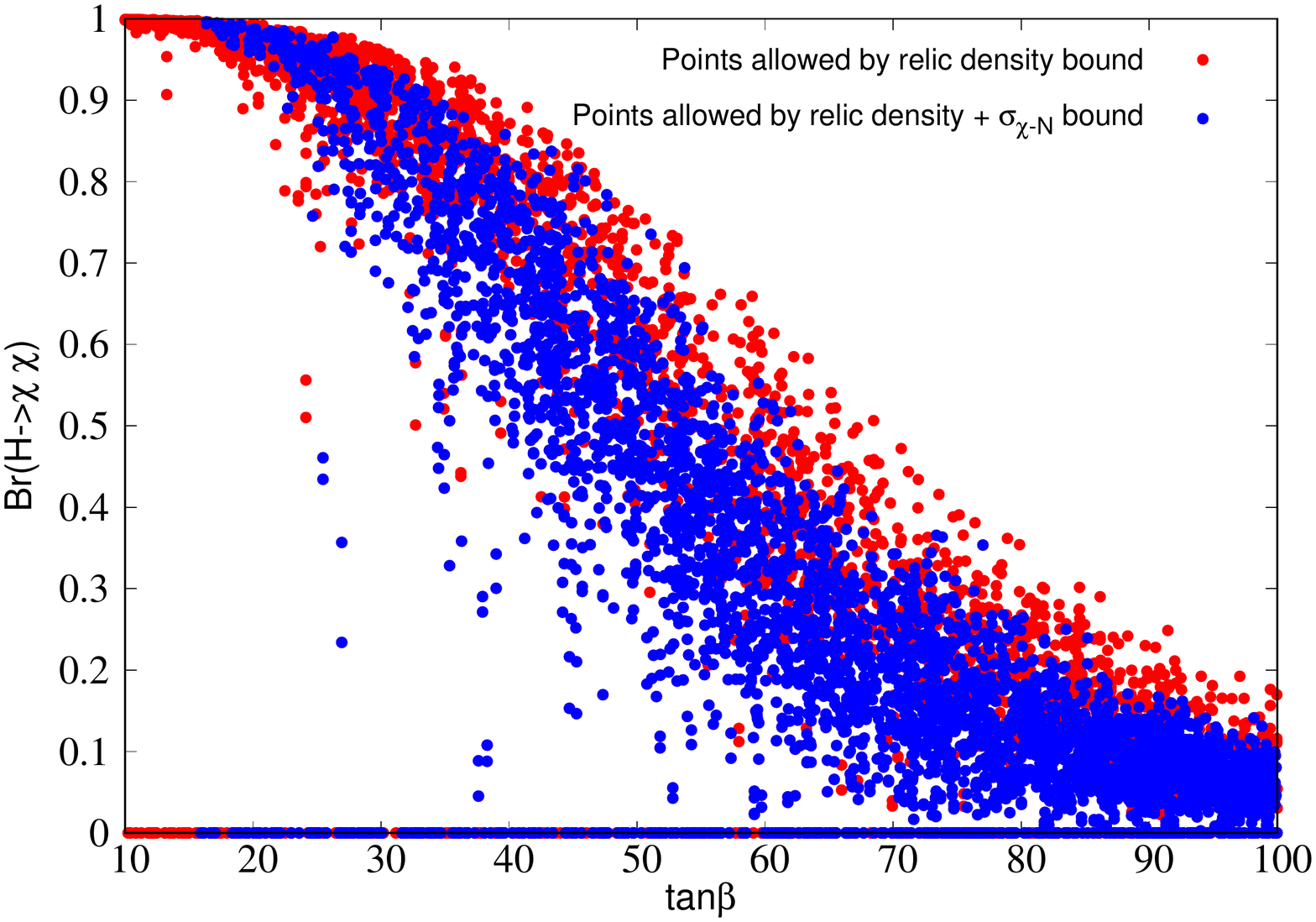}
\includegraphics[width=7.1cm, height=7cm]{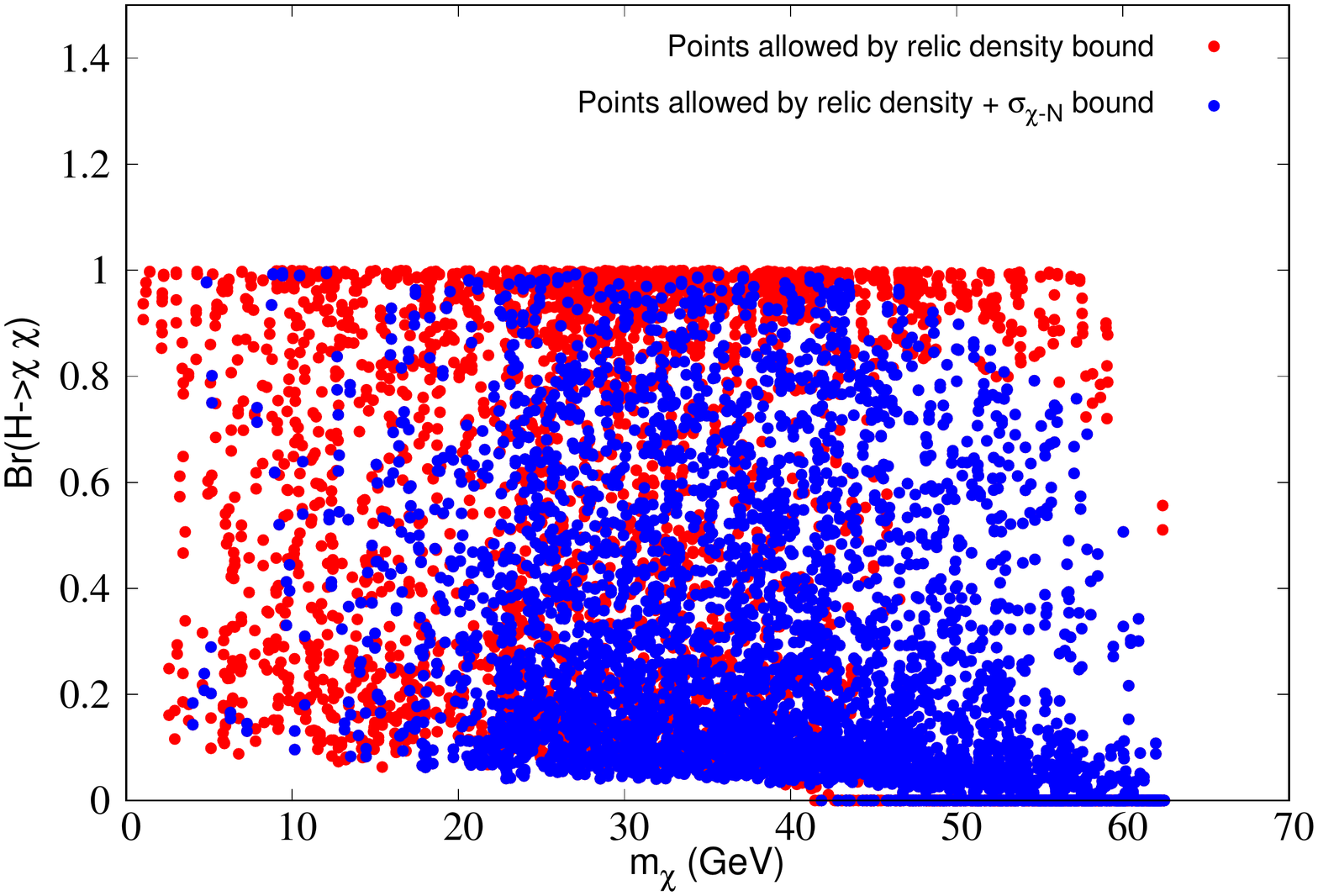}

\includegraphics[width=7.1cm, height=7cm]{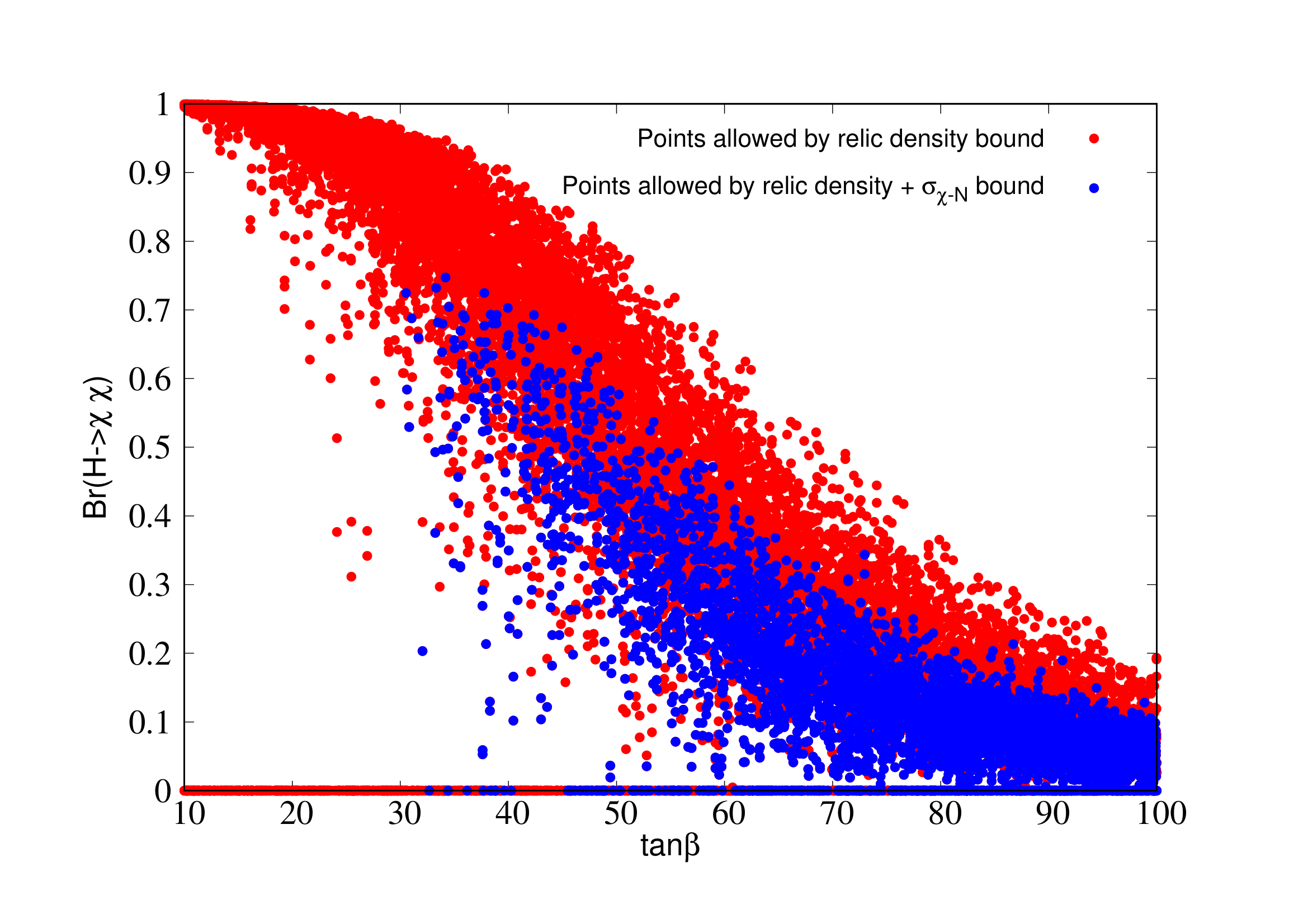}
\includegraphics[width=7.1cm, height=7cm]{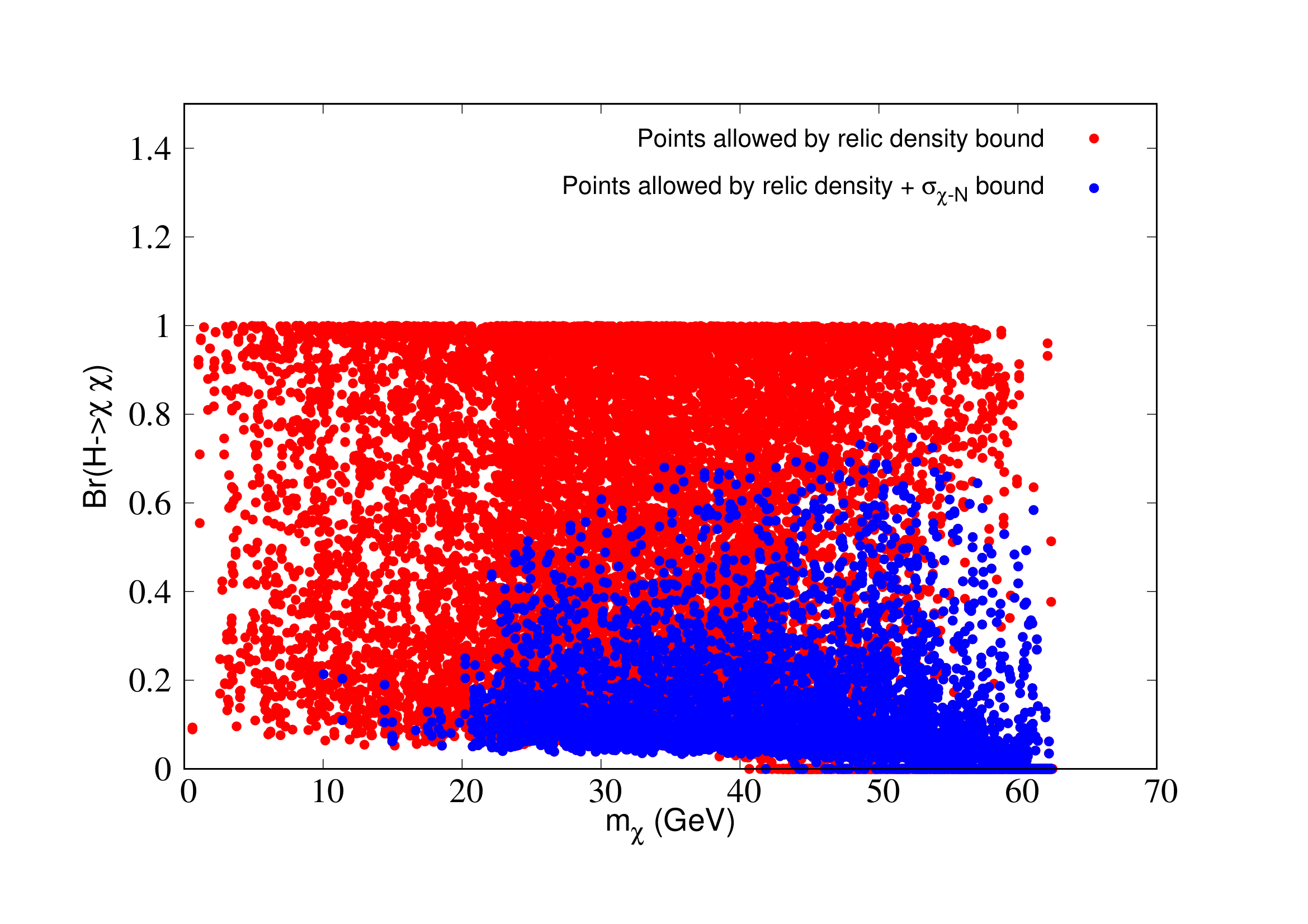}

\caption{Dependence of $BR(H \rightarrow \chi \chi)$ on $\tan \beta$(left) and $m_{\chi}$ (right) for Scenario 2. Upper two plots represent WS Yukawa case where the lower two represent RS Yukawa case.}
\label{brhxx_tb_mchi_scenario2}
\end{figure}

In Figure~\ref{brhxx_tb_mchi_scenario1} and Figure~\ref{brhxx_tb_mchi_scenario2}, we can see the dependence of BR($H \rightarrow \chi \chi)$ on $\tan \beta$ and $m_{\chi}$ where all other parameter are marginalized for Scenario 1 and Scenario 2. We can see that WS and RS cases are slightly different from each other. For Scenario 1,(Figure~\ref{brhxx_tb_mchi_scenario1}), in the RS domain we can get a very large BR($H \rightarrow \chi \chi) \approx 100\%$ in certain regions with moderate $\tan \beta$, while satisfying relic as well as direct detection constraints, which can not be attained for WS domain. In WS case, in order to satisfy XENON1T bound, one can achieve only moderate BR($H \rightarrow \chi \chi$) for $\tan \beta$ in the range $\approx$ 25 to 50. Both in WS and RS, at very large $\tan \beta$ we end up with very low BR($H \rightarrow \chi \chi)$).

On the other hand, in Scenario 2, WS and RS cases show different behavior than Scenario 1. Here we can see in Figure~\ref{brhxx_tb_mchi_scenario2}, in WS domain we can achieve a very large BR($H \rightarrow \chi \chi$) $\approx 100\%$, with moderate $\tan \beta$ but in RS cases we can achieve upto $\approx$ 70\% BR.

Next, we concentrate on dependence of BR($H \rightarrow \chi \chi$) on mass of DM particle $m_{\chi}$. We can see that for Scenario 1, to get large to moderate BR($H \rightarrow \chi \chi$) $m_{\chi}$ has to be small. As $m_H$ can be large in this case, $H \rightarrow Z A$ decay mode will open. Therefore, to increase the BR($H \rightarrow \chi \chi$), one has to increase the available phase space, ie. small $m_\chi$ will be favored. 
In Scenario 2, the mass of $H < 125.0$ GeV, available phase space for $H \rightarrow Z A$ is already small, therefore we can get a moderate to large BR($H \rightarrow \chi \chi$). However, RS case with large BR($H \rightarrow \chi \chi$) is restricted from direct detection bound as mentioned earlier.

\begin{figure}[!hptb]

\includegraphics[width=7.4cm, height=7cm]{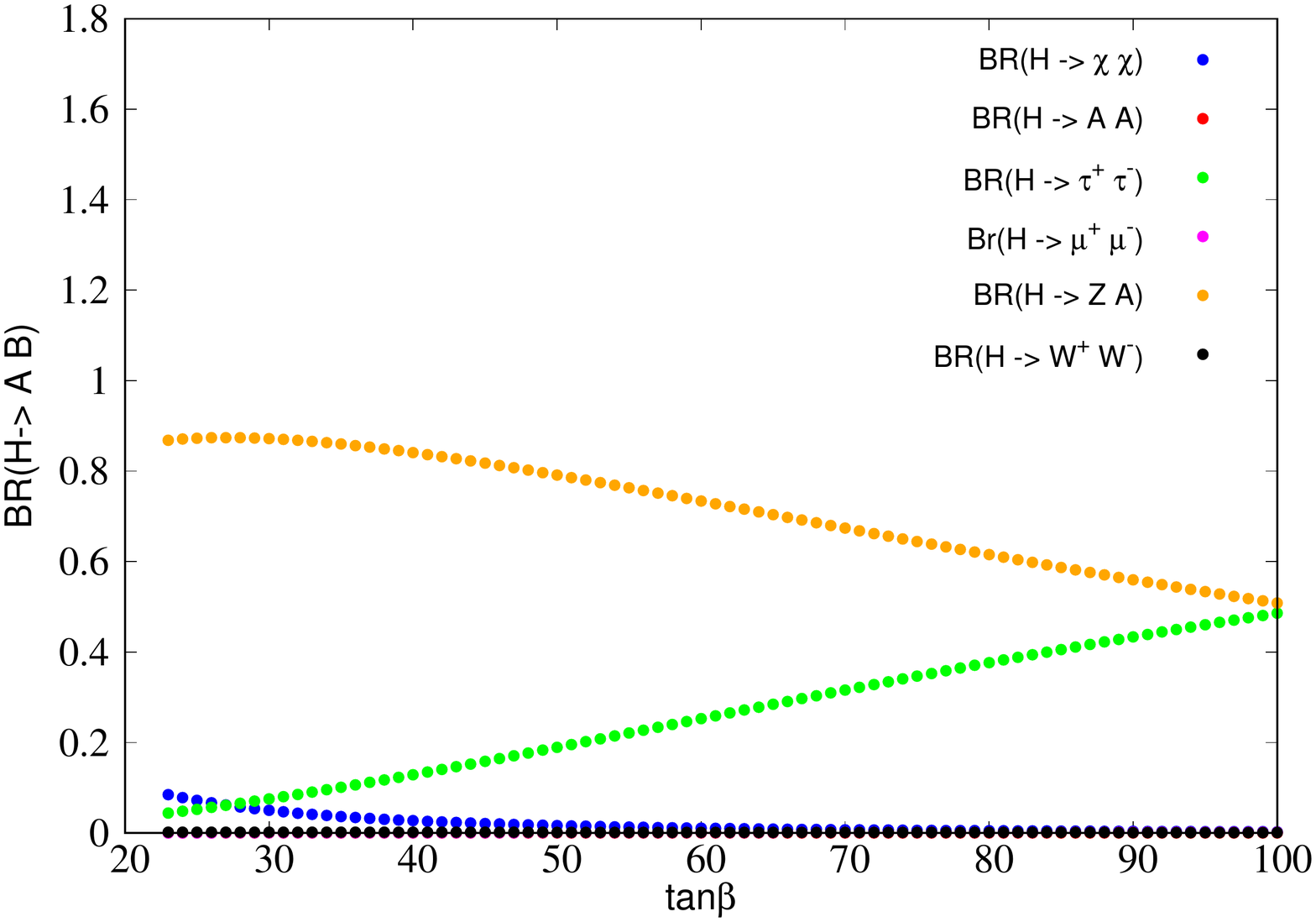}
\includegraphics[width=7.4cm, height=7cm]{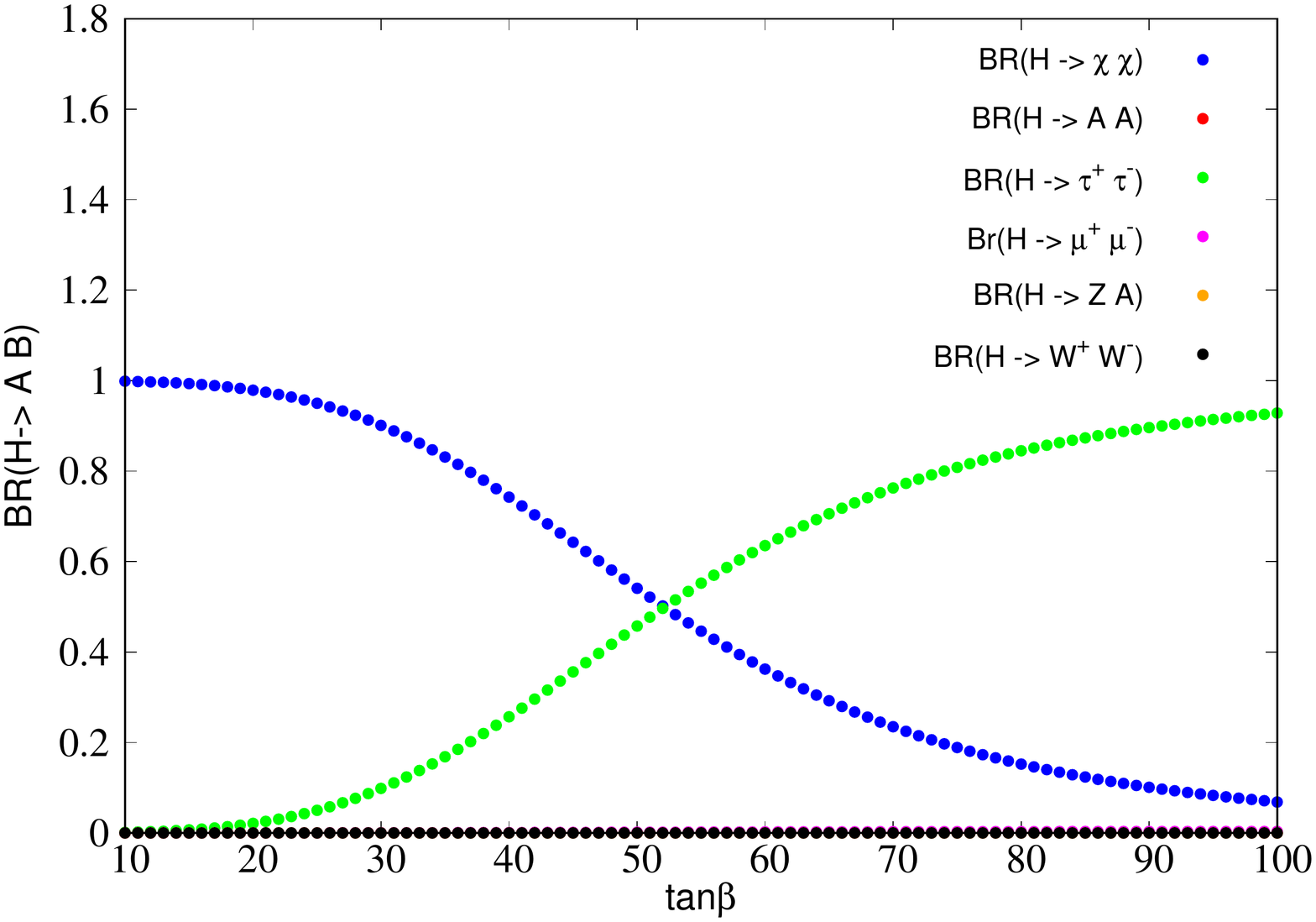}
\centering

\caption{Dependence of $BR(H \rightarrow A B)$ on $\tan \beta$ where $A$ and $B$ stand for any set of two particles for which the $Br(H \rightarrow A B)$ gives dominant contributions in the region $m_A < \frac{m_h}{2}$) for Scenario 1 (left) and Scenario 2 (right).}
\label{brhaa_tb}
\end{figure}
\begin{figure}[!hptb]

\includegraphics[width=7.2cm, height=7.0cm]{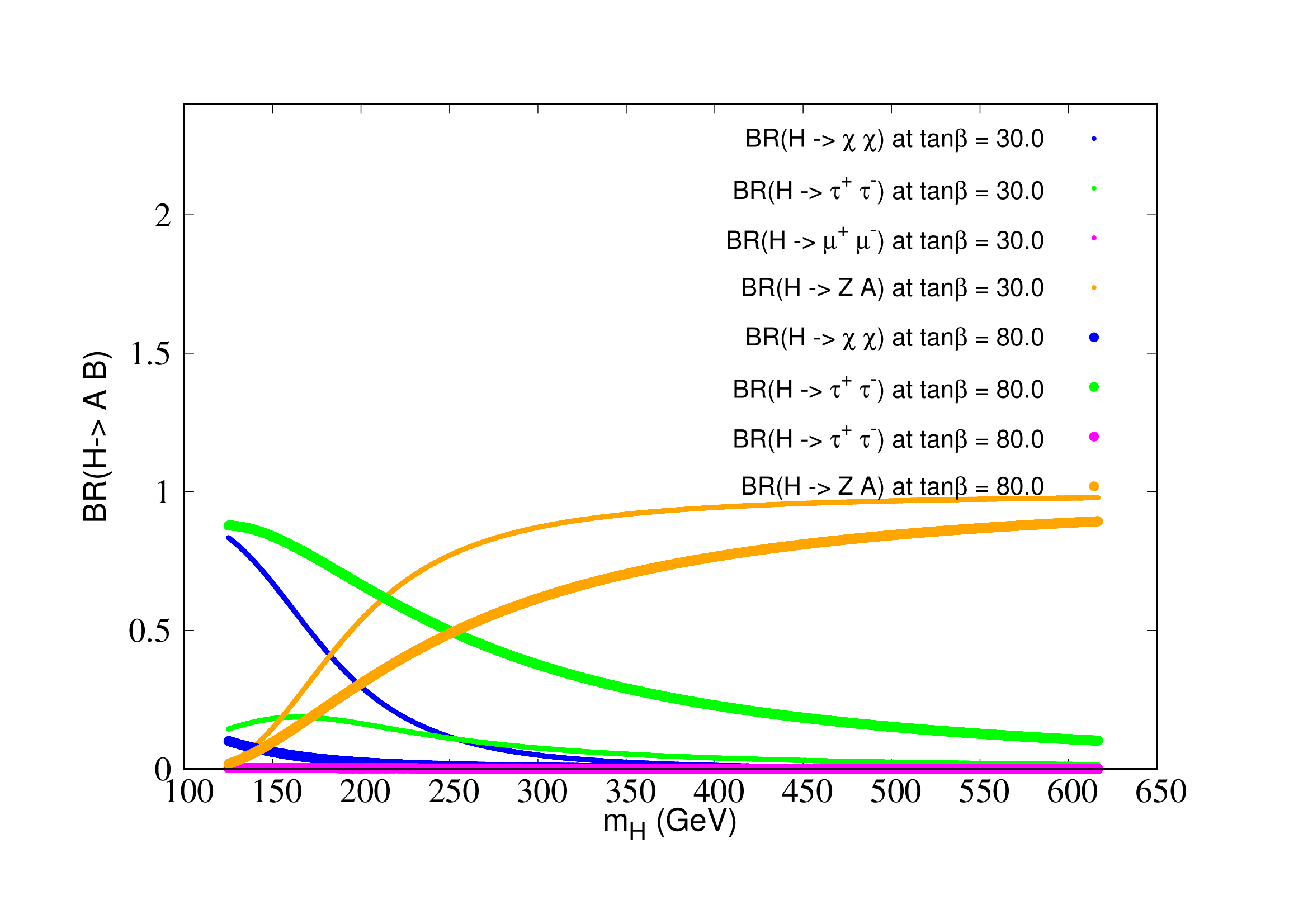}
\includegraphics[width=7.2cm, height=7.0cm]{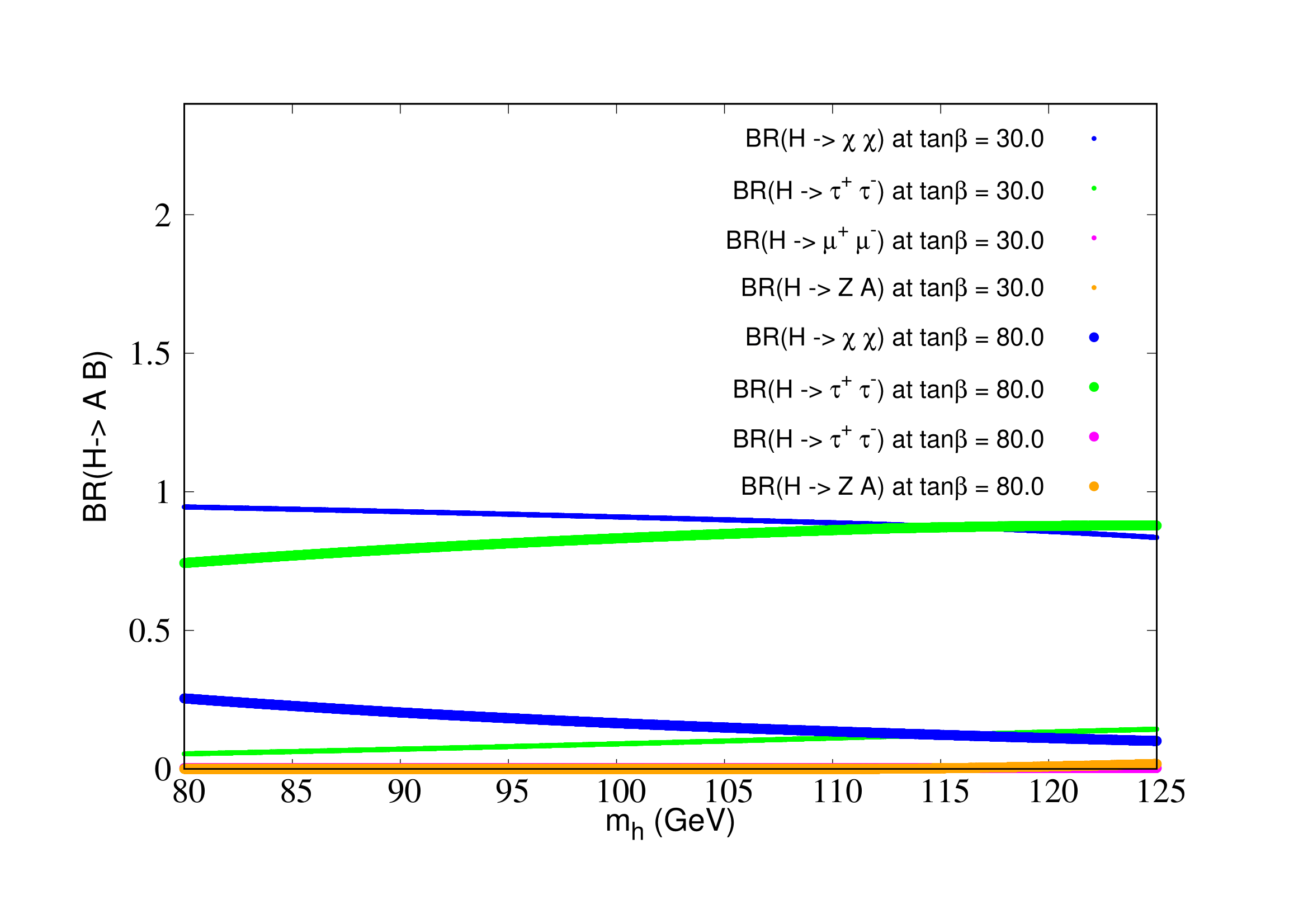}
\centering

\caption{Dependence of $BR(H \rightarrow A B)$ on $m_H$ for two different values of $\tan \beta$ where $A$ and $B$ carry similar definition as that in Figure ~\ref{brhaa_tb} in the region $m_A < \frac{m_h}{2}$ for Scenario 1 (left) and Scenario 2 (right).}
\label{brhaa_mH}
\end{figure}

In the Figure~\ref{brhaa_tb}, we can see the dependence of branchings of heavy Higgs decay in major channels with $\tan \beta$. 
For Scenario 1, we choose relevant parameters such as $\sin (\beta-\alpha)$ = 0.999, $m_H$ =  $m^{\pm}_{H}$ = 300.0 GeV, $m_\chi$ = $\frac{m_H}{2} - 10.0$ GeV, $m_A$ =20.0 GeV, $\lambda_{2s} \approx 10$. On the other hand, for Scenario 2, we choose $\cos (\beta-\alpha)$ = 0.999, $m_H$ = $m^{\pm}_{H}$ = 104.0 GeV, $m_\chi$ = $\frac{m_H}{2} - 10.0$ GeV, $m_A$ = 20.0 GeV. We can see that  for scenario 1, as $H$ can decay to $ZA$ pair, at moderate $\tan \beta$ BR($H \rightarrow ZA$) becomes the dominant channel with $>$50\% BR. However, BR($H\rightarrow ZA$) decreases with increasing $\tan \beta$ and BR($H \rightarrow \tau \tau$) starts to increase. These two decay modes share $\approx$ 50\% BR among themselves at very large $\tan \beta$. We can also see that BR($H \rightarrow \chi \chi$) decreases with $\tan\beta$. For Scenario 2, we see similar pattern. However, in this particular case, $H$ has no on-shell decay to $ZA$ final state, the contest is mostly between $H \rightarrow \tau\tau$ and  $H \rightarrow \chi\chi$ final states. The former increases with $\tan\beta$ whereas the latter decreases in large $\tan \beta$ region.

In Figure~\ref{brhaa_mH}, we can see the dependence of branching fractions on $m_H$, with two different $\tan\beta$ values. For Scenario 1, we can see that as $m_H$ increases, BR($H\rightarrow ZA$) increases and branching fractions to other two major decay channels, $\tau\tau$ and $\chi\chi$ decrease. On the other hand, for Scenario 2, as $m_H$ is almost fixed at a low value, the branching ratios in all possible channels remain more or less fixed in this region.

Considering all kind of dependencies of BR($H \rightarrow \chi \chi$), we choose six benchmark points (BP's) from different regions of parameter space. For Scenario 1 and Scenario 2 we listed them in Table~\ref{bp_gt125_ws_lt62} and Table~\ref{bp_lt125_ws_lt62} respectively.

\begin{table}[!hptb]
\begin{center}
\begin{footnotesize}
\begin{tabular}{| c | c | c | c |}
\hline
& BP1 & BP2 & BP3 \\
\hline
$m_H$ in GeV & 200.0 & 259.13 & 153.865\\
\hline
$m_A$ in GeV & 80.0 & 11.8517 & 69.0 \\
\hline
$m_{H^{\pm}}$ in GeV  & 240.89 & 257.904 & 195.0 \\
\hline
$m^2_{12}$ in GeV$^2$ & 525.965789 & 3197.1 & 347.8535\\
\hline
$m_\chi$ in GeV  & 70.0 & 100.0 & 30.5 \\
\hline
$\lambda_{2s}$ & -0.00202 & -0.02751 & -0.00272\\
\hline
$\lambda_{1s}$ & 11.574 & 12.0275 & 12.5027\\
\hline
$\tan \beta$ & 76.0 & 20.9107 & 68.0\\
\hline
$\sin(\beta - \alpha)$ & 0.9996 & 0.993964 & 0.999996\\
\hline
$y_h^{\ell} \times \sin(\beta - \alpha)$ & -1.14932972 & -1.292301198257 & 0.8076599172\\
\hline
$Br(H\rightarrow \chi \chi)$ & 6.734168\% & 32.3\% & 24.8\%\\
\hline
$Br(H^{\pm}\rightarrow \tau \nu_{\tau})$ & 59.87\% & 5.33\% & 73.89\%\\
\hline
$\Omega h^2$ & 7.24$\times 10^{-4}$ & 1.16$\times 10^{-6}$ & 6.71$\times 10^{-4}$\\
\hline
$\sigma_{\chi - N}$ in $pb$ & 8.214$\times 10^{-11}$ & 1.038$\times 10^{-8}$ & 1.396$\times 10^{-10}$\\
\hline
\end{tabular}
\end{footnotesize}
\caption{\it Benchmark points for Scenario 1.}
\label{bp_gt125_ws_lt62}
\end{center}
\end{table}

For Scenario 1, we choose three benchmarks (see Table~\ref{bp_gt125_ws_lt62}). BP1 represents WS Yukawa with $m_A > \frac{m_h}{2}$, while BP2 has WS Yukawa with $m_A < \frac{m_h}{2}$. BP3 represents RS Yukawa with $m_A > \frac{m_h}{2}$. The justification behind such choices, come from the theoretical, and collider constraints along with the observed $g_\mu - 2$ anomaly. Notably, observed muon anomaly favors large $\tan \beta$ especially for $m_A > \frac{m_h}{2}$. Large $\tan\beta$ region implies large BR($H\rightarrow\tau\tau$) and reduced BR($H \rightarrow \chi \chi$). Therefore as discussed earlier, for this scenario, in WS cases, we can not get substantial BR in $H \rightarrow \chi\chi$ channel. From Figure~\ref{brhxx_tb_mchi_scenario1} we can see that at $\tan \beta$ close to 76(BP1) and 21(BP2), we can only have low BR if we have to satisfy all constraints. On the other hand, for RS cases we can in principle have comparatively large BR. However, due to large $\tan\beta$ in case of BP3, BR($H \rightarrow \chi \chi) \approx 20\%$.

\begin{table}[!hptb]
\begin{center}
\begin{footnotesize}
\begin{tabular}{| c | c | c | c |}
\hline
& BP4 & BP5 & BP6 \\
\hline
$m_H$ in GeV & 117.409 & 93.6073 & 121.446\\
\hline
$m_A$ in GeV & 67.0 & 25.0 & 64.0 \\
\hline
$m_{H^{\pm}}$ in GeV  & 167.0 & 135.0 & 171.0 \\
\hline
$m^2_{12}$ in GeV$^2$ & 196.796761 & 282.655697 & 216.79898\\
\hline
$m_\chi$   & 40.0 & 6.0 & 50.5 \\
\hline
$\lambda_{2s}$ & -0.00254 & -0.00396 & -0.00266\\
\hline
$\lambda_{1s}$ & 12.3745 & 3.804 & 12.3027\\
\hline
$\tan \beta$ & 70.0 & 31.0 & 68.0\\
\hline
$\sin(\beta - \alpha)$ & -0.00141421 & 0.00447212477 & -0.0316188235075\\
\hline
$y_h^{\ell} \times \sin(\beta - \alpha)$ & 0.901003399& 1.138614482 & -1.14940501\\
\hline
$Br(H\rightarrow \chi \chi)$ & 28.2\% & 67.3\% & 24.0\%\\
\hline
$Br(H^{\pm}\rightarrow \tau \nu_{\tau})$ & 90.89\% & 63.41\% & 86.21\%\\
\hline
$\Omega h^2$ & 9.59$\times 10^{-5}$ & 1.88$\times 10^{-3}$ & 3.78$\times 10^{-5}$\\
\hline
$\sigma_{\chi - N}$ in $pb$ & 1.704$\times 10^{-10}$ & 4.106$\times 10^{-8}$ & 8.351$\times 10^{-11}$\\
\hline
\end{tabular}
\end{footnotesize}
\caption{\it Benchmark points for Scenario 2.}
\label{bp_lt125_ws_lt62}
\end{center}
\end{table}

Similarly for Scenario 2, three benchmarks are chosen(see Table~\ref{bp_lt125_ws_lt62}). We choose BP4 with RS Yukawa and $m_A > \frac{m_h}{2}$, BP5 with RS Yukawa and $m_A < \frac{m_h}{2}$ and BP6 with WS Yukawa and $m_A > \frac{m_h}{2}$. Here too, our choices are guided by all relevant constraints along with the observed $g_\mu - 2$.
As the observed $g_{\mu}-2$ favors large $\tan \beta$ for $m_A > \frac{m_h}{2}$, BR($H \rightarrow \chi \chi$) is on the lower side as discussed earlier. However, in Scenario 2 $m_H$ is comparatively smaller and therefore $H \rightarrow Z A$ channel is absent. Therefore, one can achieve considerable BR($H \rightarrow \chi \chi$). From Figure~\ref{brhxx_tb_mchi_scenario2}, we can see that in RS case at $\tan \beta \approx 70$(BP4) we get BR $\approx$ 28\%, while for $\tan \beta \approx 31$(BP5), BR($H \rightarrow \chi \chi) \approx 70\%$ is achieved. For WS case, with $\tan \beta \approx 68$(BP6), one gets $\approx$ 19\% BR in $H\rightarrow \chi\chi$ channel.

\section{Collider Analysis : Cut-based}
\label{sec5}

Having discussed all important aspects of our model and chosen our benchmarks, we proceed to perform a detailed collider analysis, in terms of probing such a scenario at the high luminosity runs of LHC. The prospect at collider experiments relies heavily on choosing the signal, ie. a suitable production mechanism as well as appropriate final states. The production of DM particle $\chi$ will result in significant $\slashed{E_T}$ in the final state. However in order to detect that, one essentially has to consider some visible particle in the final state which will recoil against the DM. As DM in our case couples significantly only with $H$, we will focus on the scenario where DM pair is produced from the decay of $H$. However, we discussed earlier, the non-standard scalars in our model couple very weakly with the quarks. Therefore, the best-suited production mechanism for $H$ in this case will be via Drell-Yan process. We will focus on the following channel in our analysis.

\medskip
$p p \rightarrow H^{\pm}H ( H^{\pm} \rightarrow \tau_{had} + \nu_{\tau}, H\rightarrow \chi\chi$ )

\medskip
\noindent
As $H^{\pm}$ is leptophilic in nature, especially at large $\tan\beta$, it decays into $\tau\nu_{\tau}$ with significant branching ratio($\gsim$ 60\%) in relevant parts of the parameter space. We have considered in this work the hadronic decay mode of $\tau$(BR$\approx 65\%)$.

\subsection{Signal and backgrounds}

The signal and dominant backgrounds for our process will be discussed in detail shortly. All the signal and the backgrounds are calculated at
next-to-leading order, using Madgraph@MCNLO~\cite{Alwall:2014hca}.
For multi-jet events, MLM matching has been performed with appropriate XQCUT.
We have used nn23lo1 parton distribution functions. We perform showering and hadronization via PYTHIA8~\cite{Sjostrand:2006za}.
Delphes-3.4.1~\cite{deFavereau:2013fsa} has been used for detector simulation, with in-built Fastjet~\cite{Cacciari:2006sm} implementation for jet formation. We mention here that at HL-LHC, due to large increase in instantaneous luminosity, pile-up effects are expected to be significant. We have included pile-up in our analysis, assuming 140 average pile-up interactions.

\bigskip
\noindent
{\bf Signal:}

\medskip  
\noindent
Our sought after signal will be one hard $\tau$-tagged jet(hadronic $\tau$ or $\tau_h$) + $\slashed{E_T}$. We also put a veto on $b$-jets and hard leptons($p_T^{\ell} > 10$GeV). Such final states have been searched by CMS and ATLAS in the past~\cite{ATLAS:2014otc,CMS:2019bfg}. We closely follow the analysis strategy of ~\cite{CMS:2019bfg} in this work.

\bigskip
\noindent
{\bf Backgrounds:}
\medskip

\noindent
{\it QCD jets faking as $\tau$ jets}: The leading background for this channel comes from QCD multi-jet events. In this case, at least one of the light QCD jets are misidentified as a $\tau$-jet. Mistag rate in this case is taken to be $\approx 1\%$. The large cross-section of QCD multijet process makes the tail of this background significant. However, as there is no real source of $\slashed{E_T}$ in this case, the $\slashed{E_T}$ can arise from jet-energy mismeasurement. Suitable cuts are to be applied to reduce this background. We put a generator level cut of 50 GeV on two leading jets in order to gain statistics.

\medskip 
\noindent
{\it $W$+jets}: The next major background comes from $W$+jets events. In this case there are two possibilities. Firstly, $W$ can decay into $\tau_h+\nu_\tau$ final state and thus constitute an irreducible background. On the other hand, it is also possible to get misidentified $\tau$-jets from leptons of $W$-decay. However, the rate of lepton mistagging as $\tau$ is really small.  Furthermore, there is real source of $\slashed{E_T}$ due to neutrinos from $W$-decay. In addition, large production cross-section of $W$+jets background makes it challenging.

\medskip
\noindent
{\it $t \bar t$}: Another crucial background with real source of $\slashed{E_T}$ is $t\bar t$. Here too, $\tau_h$ can occur from the decay of $W$ from $t \bar t$. In addition, light-jets from top decay can also be misidentified as $\tau_h$. However, the aforementioned $b$-veto and lepton-veto on the events help us control the background to a large extent.   

\medskip

\noindent
{\it $WZ$+jets}: This irreducible background has larger source of $\slashed{E_T}$ compared to $W$+jets background, when $Z$ decays into a pair of neutrinos and the possibility of getting real as well as mistagged $\tau_h$ in the final state. However, its cross-section is much smaller compared to the previously mentioned backgrounds and therefore it plays only a sub-leading role in our analysis.

\medskip
\noindent
{\it Single top}: Lastly, single top production can also act as a possible source of background with real source of $\slashed{E_T}$ and real or fake $\tau_h$ from top decay. But due to small cross-section, this also acts as a sub-dominant background.

\subsection{Distributions}

In order to suppress the aforementioned backgrounds, one is motivated to look into various kinematic observables, which can distinguish between signal and backgrounds efficiently. Thereafter, suitable cuts are to be applied on those variables. With this agenda, we first show the distributions for all signal BP's as well as the three major backgrounds.

\begin{figure}[!hptb]
	\centering
         \includegraphics[width=7.5cm,height=6.5cm]{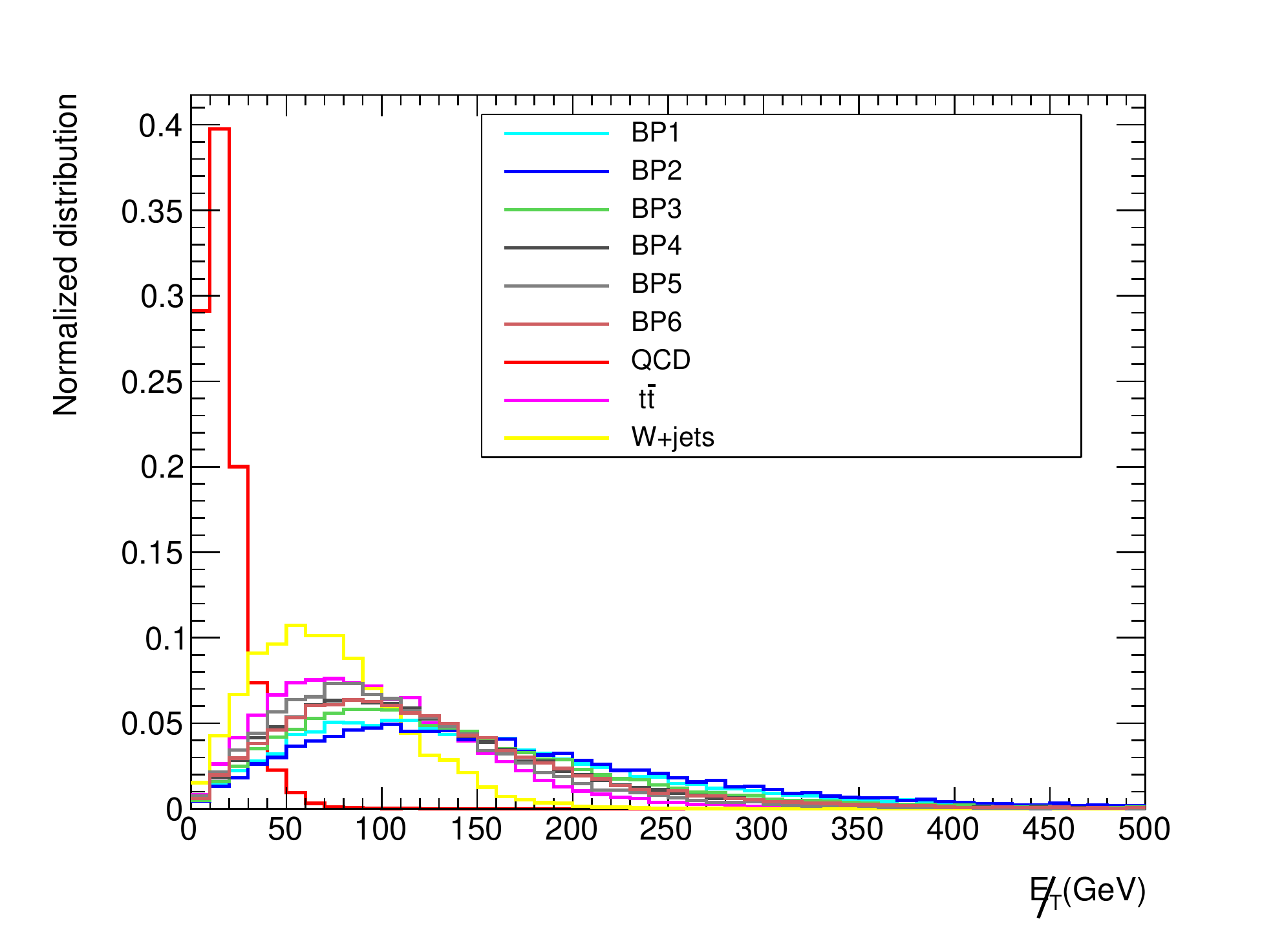}
\includegraphics[width=7.5cm,height=6.5cm]{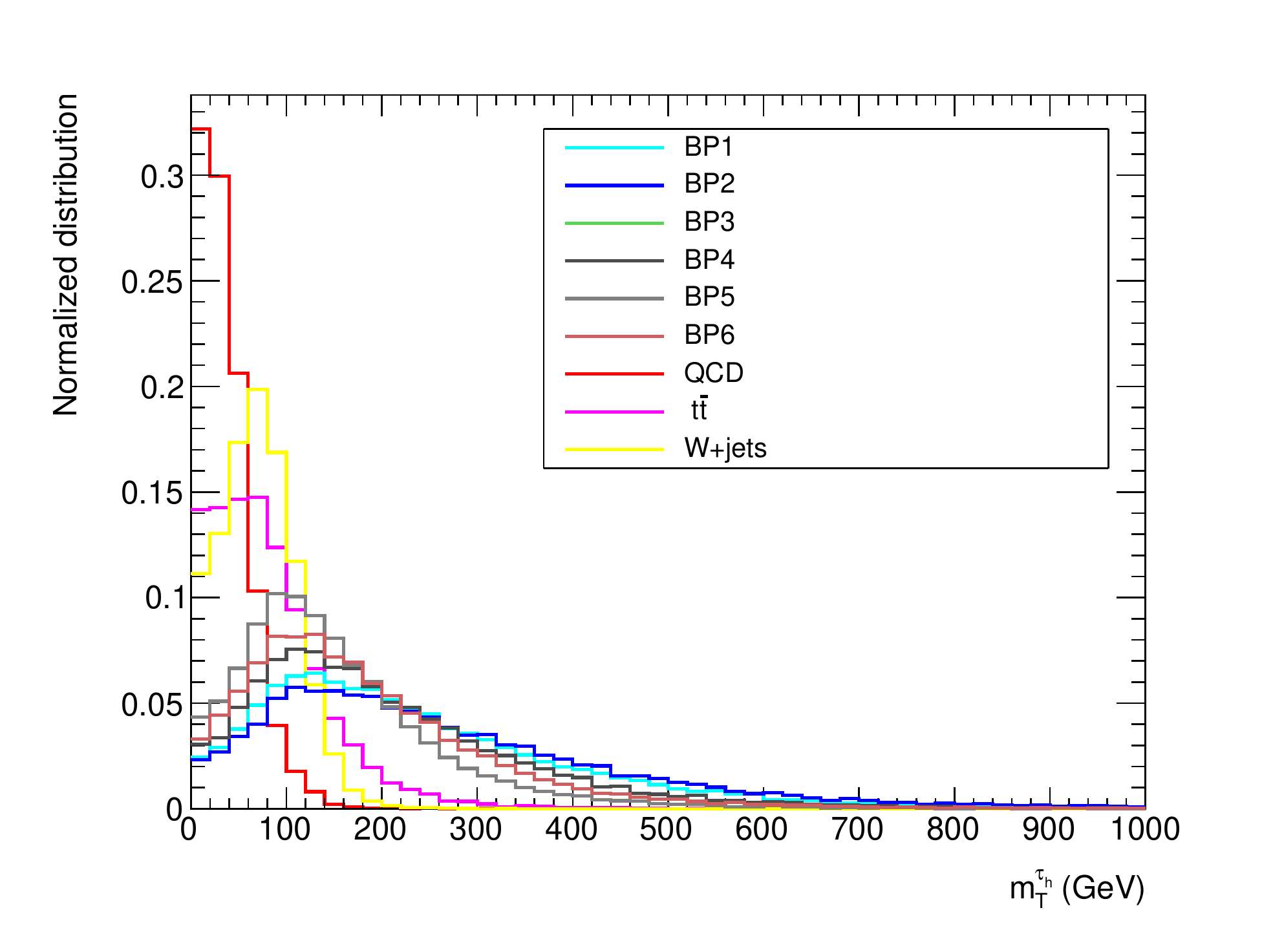} \\
  \includegraphics[width=7.5cm,height=6.5cm]{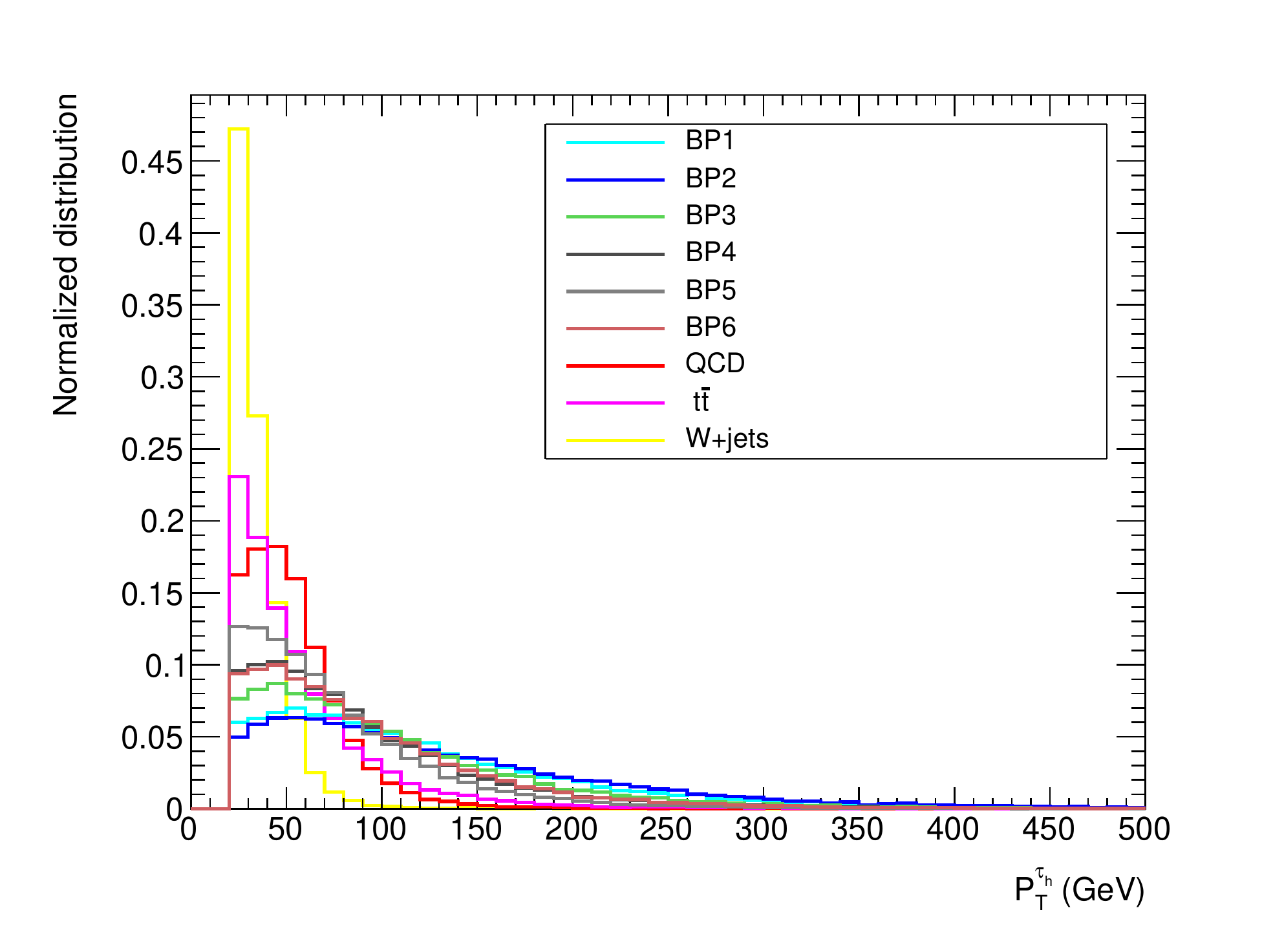}
         \includegraphics[width=7.5cm,height=6.5cm]{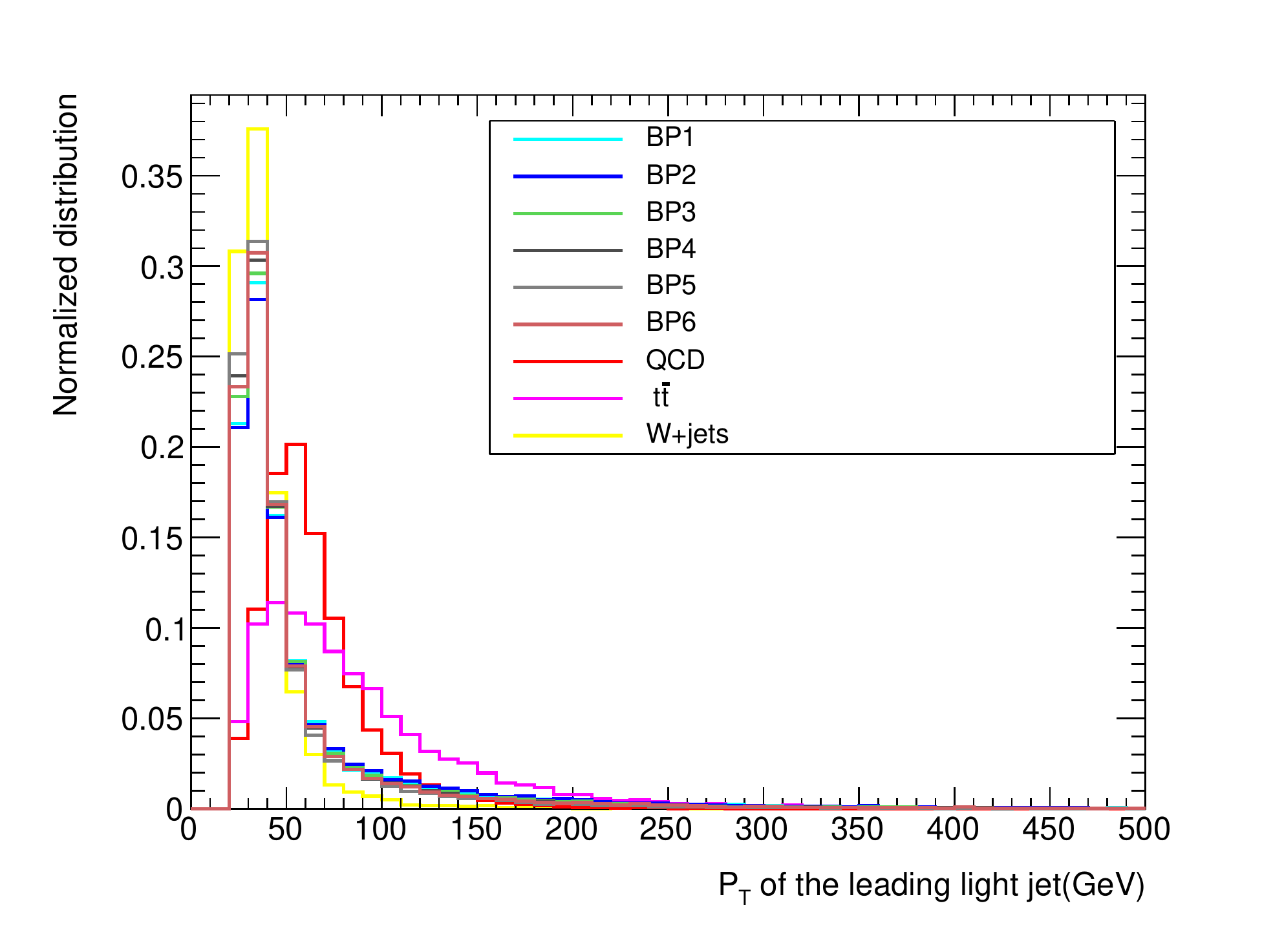} \\
\includegraphics[width=7.5cm,height=6.5cm]{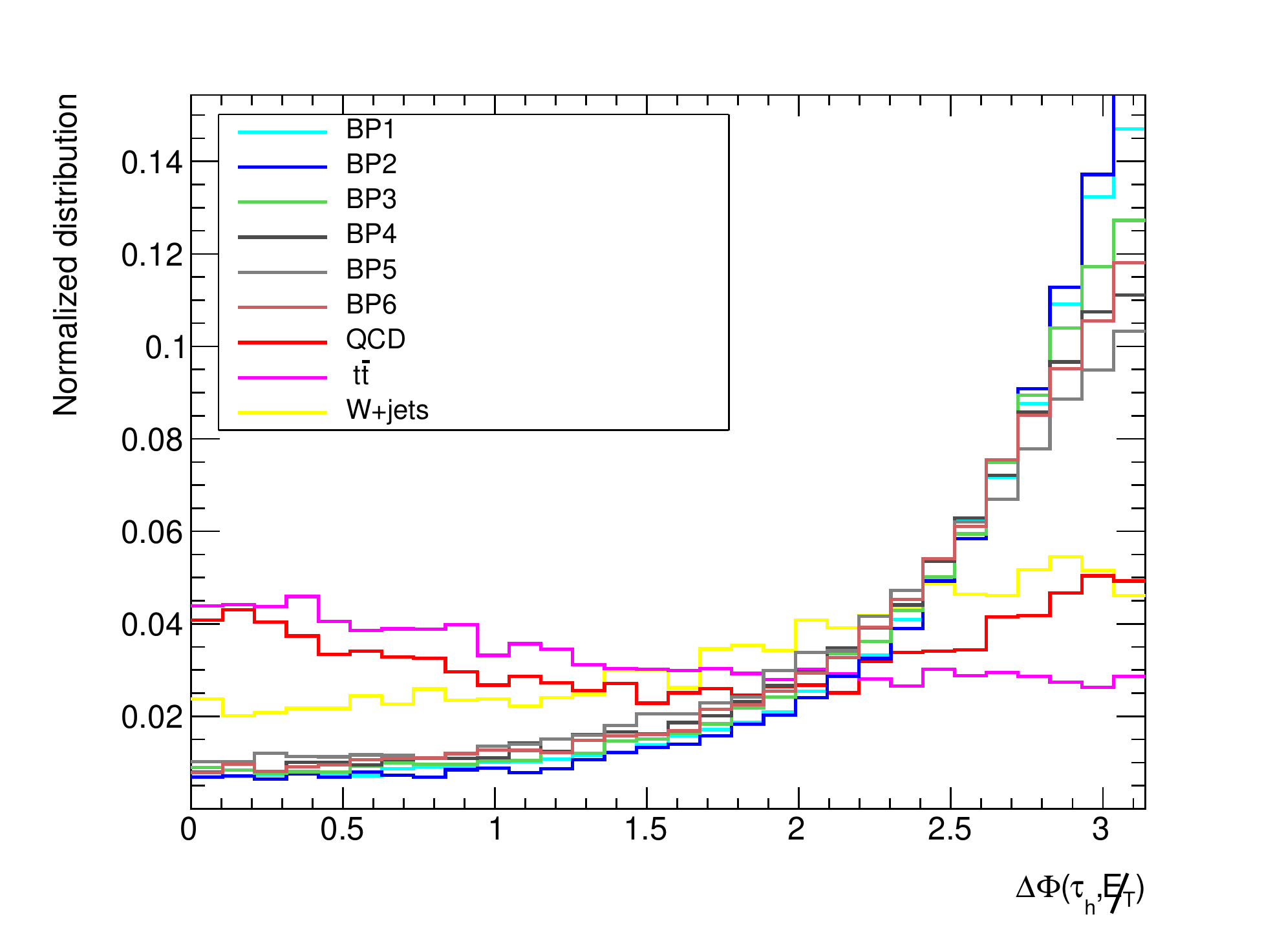} 
  \includegraphics[width=7.5cm,height=6.5cm]{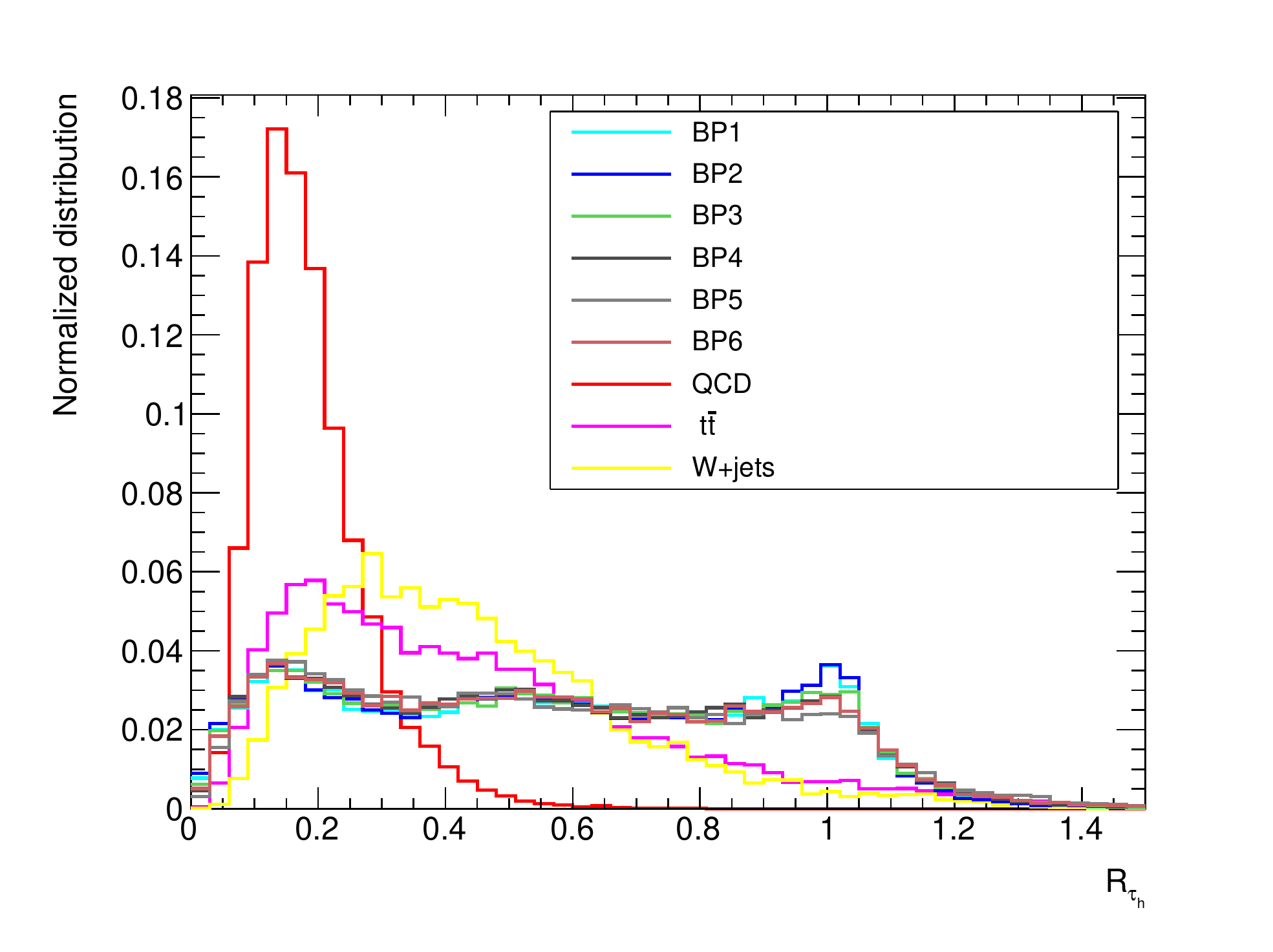}
	\caption{Normalized distribution for signal as well as backgrounds.}
	\label{dist}
\end{figure}

In Figure~\ref{dist}(top left), we show the $\slashed{E_T}$ distribution. QCD background has no real $\slashed{E_T}$, and the only source of $\slashed{E_T}$ is through jet energy mismeasurement. We can see that in case of QCD multijet, $\slashed{E_T}$ peaks on the lower side. On the other hand, $W$+jets and $t \bar t$ show significant $\slashed{E_T}$. However, the signal BP's lie further on the higher side of $\slashed{E_T}$ spectrum, since in that case $\slashed{E_T}$ comes from the decay of a heavy scalar $H$. It is indeed clear that a strong $\slashed{E_T}$ cut is required to reduce the backgrounds.

Figure~\ref{dist}(top right) shows the transverse mass distribution($m_T^{\tau_h}$) in the hadronic final state. The $m_T^{\tau_h}$ observable is defined as follows.

      \begin{equation}
          m_T^{\tau_h} = \sqrt{2 p_T(\tau_h)\slashed{E_T} (1 - \cos \Delta \phi(\vec{\tau_h},\vec{\slashed{E_T}}})
       \end{equation}

\noindent
In case of $W$+jets and $t\bar t$(semileptonic) backgrounds, the $\tau_h$ is coming from the decay of $W$. Therefore in such scenarios, the $m_T^{\tau_h}$ distribution will be restricted by the $W$ mass. For signals on the other hand, $\tau_h$ comes from the decay of heavy $H^{\pm}$ and therefore $m_T^{\tau_h}$ distribution for signal BP's are shifted towards higher values. This observable also plays an important role to distinguish between the signal and backgrounds.

Figure~\ref{dist}(centre left), we show the $p_T$ of the tau-tagged jet. The signal distributions for $p_T$ of $\tau_h$, tend towards higher values because $\tau$ from the decay of $H^{\pm}$ decay are boosted compared to those from $W$ decay. We should mention here that the QCD jets are generated with a 50 GeV cut and its distribution therefore peaks at a higher value. In our analysis, we demand a hard $\tau$-tagged jet to enhance signal over backgrounds.

Figure~\ref{dist}(centre right) shows the $p_T$ distribution of the leading light-jet. In case of QCD or $t \bar t$ background, the leading light-jet is generated in hard process and hence has similar $p_T$ as that of the $\tau$-tagged jet. On the other hand, in case of signals, hard $\tau_h$ recoils again the DM pair and light-jets are produced mostly via showering. The $p_T$ of the leading light-jet in this case is rather soft. Therefore, an upper cut-off on leading light-jet $p_T$ will help us distinguish between signal and QCD or $t\bar t$ backgrounds. One can see in case of $W$+jets background, the distribution is similar to the signal case, since here too, the light-jets are produced in showering and are  on the softer side.

In Figure~\ref{dist}(bottom left), we show the $\Delta\phi$ distribution between the direction of $\slashed{E_T}$ and the $p_T$ direction of $\tau_h$. We can see that, in case of QCD background, the $\slashed{E_T}$ is aligned with either of the jets, and the distribution also points towards that. In case of $t\bar t$ background, $\Delta\phi$ shows a somewhat flat distribution. $W$+jets distribution shows a peak at $\pi$. However, $\tau_h$ and $\slashed{E_T}$ there are not exactly back to back. One can see that for signals they are mostly back to back. A $\Delta\phi$ cut can be applied to make an effective signal-background separation.

Lastly, in Figure~\ref{dist}(bottom right), we present an observable $R_{\tau_h}$. This observable is sensitive to the polarization of $\tau$. The $\tau$-lepton produced from $W$ decay will be left-handed and the ones produced from $H^{\pm}$ decay will be right-handed from angular momentum conservation. $\tau$ decays to a charged pion $\pi^+$ and a neutral pion $\pi^0$ and a neutrino, via intermediate $\rho$ meson with $\approx$25\% BR. The angular distributions of the $\tau$ decay products are extremely sensitive to $\tau$ polarization. The angular observables can be translated into energy observables and can be utilized to distinguish between left-and right-handed $\tau$ in the lab frame. Various such observables  been pointed out in~\cite{ATLAS:2012tqa,CMS:2019bfg}. In this work, we consider one such observable called $p_T$-imbalance $R_{\tau_h}$ defined as follows.

\begin{equation}
R_{\tau_h} = \frac{p_T^{track}}{p_T^{\tau_h}}
\end{equation}  

\noindent
For signal with right-polarized $\tau$, the distribution peaks at a larger value compared to the left-handed($W$+jets and $t\bar t$) or unpolarized $\tau_h$(QCD fakes). This observable certainly helps us distinguish between our signal and various backgrounds.

\subsection{Results}

Having discussed the kinematic observables, we proceed to find a suitable cut-flow, that 
will significantly discriminate between signal and background.

\noindent
{\bf Cut flow:}

\noindent
Pre-selection criteria : One hard $\tau$-tagged jet, 0 $b$-jet and 0 hard lepton \\
Cut1 : $p_T$ of $\tau$-tagged jet $> 50$GeV \\
Cut2 : $\slashed{E_T} > 150$ GeV \\
Cut3 : $m_T^{\tau_h} > 200$ GeV \\
Cut4 : $p_T$ of the leading light-jet $< 80$GeV \\
Cut5 : $R_{\tau_h} > 0.7$ \\

\begin{table}[!hptb]
\scriptsize{
\begin{tabular}{| c | c | c | c | c | c | c | c |}
\hline
Datasets & Cross-section& Pre-selection cut & Cut1 & Cut2 & Cut3 & Cut4 & Cut5 \\
\hline
BP1  & 0.5 fb & 685 & 540 & 263 &  243 & 194  & 72  \\
\hline
BP2  & 0.1 fb & 140 & 112 & 60 & 56 &  44 & 17  \\
\hline
BP3  & 5.3 fb & 5338 & 3889 & 1618 &  1482 & 1198 & 596 \\
\hline
BP4  & 16 fb &12773 & 8645 & 3056 & 2700 & 2195  & 1296  \\
\hline
BP5  & 56.5 fb &66598 & 39947 & 11706 & 9946 & 8349  &  2980  \\
\hline
BP6  & 11.3 fb &14207 & 9706 & 3559 & 3160  & 2584  &  934 \\
\hline
QCD multijet & $10^7$ pb & 3.8$\times 10^{11}$ & 1.6$\times 10^{11}$ & 1.8$\times 10^7$ &  6$\times 10^6$  & 4.0$\times 10^6$  & 300000 \\
\hline
$W$+jets & $10^4$ pb & 4.9$\times 10^9$ & 4.3$\times 10^8$  & 2.7$\times 10^7$ &  7.4$\times 10^6$  & 6.8$\times 10^6$  & 109188 \\
\hline
$t \bar t$ & 106 pb & 1.9$\times 10^7$  & 7.5$\times 10^6$  & 1.2$\times 10^6$ & 279840 & 133560  & 19080  \\
\hline
$WZ$+jets & 309 fb & 219866 & 51467 & 9029 &  4931 &  3856 & 445   \\
\hline
Single top & 429 fb & 96474  & 24195  & 2445 & 566 & 360  & 26  \\
\hline
\end{tabular}
\caption{Signal and background events surviving after applying the pre-selection criteria as well as selection cuts at $\sqrt s = 14$ TeV and ${\cal L} = 3000 fb^{-1}$.} 
\label{tablecutflow14}
}
\end{table}

The significance~\cite{Cowan:2010js} has been calculated using the following formula.
\begin{equation}
{\cal{S}}=\sqrt{2[(S+B) Log(1+ \frac{S}{B}) -S]}
\end{equation}

\begin{table}[!hptb]
\begin{center}
\begin{tabular}{| c | c | }
\hline
BP1  &  0.11$\sigma$  \\ 
\hline
BP2  &  0.03$\sigma$   \\
\hline 
BP3  &  0.9$\sigma$   \\
\hline
BP4  &  2.0$\sigma$   \\
\hline
BP5  &  4.5$\sigma$   \\
\hline
BP6  &  1.5$\sigma$   \\ 
\hline
\end{tabular}

\caption{\it Signal significance for the benchmark points at 14 TeV LHC with 3000$fb^{-1}$ with cut-based analysis. }
\label{significance}
\end{center}
\end{table}

\noindent
One can see that BP1 and BP2 have extremely low significance. BP1, with large $\tan \beta$ suffers due to small branching ratio. BP2 on the other hand, due to moderate $\tan\beta$, can have larger BR($H\rightarrow \chi\chi)$. But in this case $m_A$ is sufficiently small, so that $H^{\pm} \rightarrow W^{\pm}A$ channel opens up with large branching and $H^{\pm} \rightarrow \tau_h\nu$ channel suffers from a low branching ratio. BP3 is similar to BP1, except for the masses of the non-standard scalars and DM. Smaller masses enhance production cross-section as well as BR($H\rightarrow \chi\chi$) and therefore yield better significance. Benchmarks in Scenario 2, have significantly higher cross-section, due to small masses of the non-standard scalars. They also get significant BR($H\rightarrow\chi\chi$) as well as BR($H^{\pm} \rightarrow \tau\nu$), since many other possible decay modes are kinematically disallowed. BP5 fares best among all the benchmarks, due to its small masses. One can infer that, in general, Scenario 2 is much easier to probe at the collider.

\section{ANN Analysis}
\label{sec6}

After the cut-based analysis is done, we explore the possibility of improvements of our analysis with ANN~\cite{Teodorescu:2008zzb}. ANN has shown marked improvement compared to results pertaining to cut-based analyses~\cite{Hultqvist:1995ibm,Field:1996rw,Bakhet:2015uca,Dey:2019lyr,Lasocha:2020ctd,Dey:2020tfq,Bhowmik:2020spw}. In the present analysis signal yield is poor in general because of multiple reasons discussed above. Therefore, an efficient signal-background separation method has to be adopted. With this agenda, we use ANN and calculate the significance that can be achieved at HL-LHC. We use a python-based deep-learning library Keras~\cite{keras} for the analysis.

From our earlier discussion, we have identified various observables which are chosen to be the input variables or feature variables for the network. Apart from the observables discussed previously, we consider a few new observables as well. These observables were not used in the cut-based analysis, due to weak discriminating power or correlation with other observables. However, we have used them for ANN analysis. We define two such additional observables below.

\begin{equation}
  R_{bb}^{min} = min\left(\sqrt{(\pi-\Delta\phi(\tau_h,\vec{\slashed{E_T}})^2) + \Delta\phi(jet_n, \vec{\slashed{E_T}})^2 }\right)
\label{rbbmin}
\end{equation}

In Equation~\ref{rbbmin}, the minimization is done with respect to three hardest light jets in the final state~\cite{CMS:2019bfg}. This observable is used typically to reduce the QCD background where one light-jet is mistagged as $\tau_h$ and $\slashed{E_T}$ is associated with mismeasurement of jet energy. This variable does not change our result significantly while used in rectangular cut-based analysis. However, we use this observable in ANN analysis.

Another such observable is $p_\theta$~\cite{ATLAS:2012tqa}. It is a variable sensitive to $\tau$-polarization and is defined as follow.

\begin{equation}
p_\theta = \frac{E_T^{\pi^-} - E_T^{\pi^0}}{p_T^{\tau_h}}
\label{ptheta}
\end{equation}

The $p_\theta$ observable in Equation~\ref{ptheta} is expected to be correlated with $R_{\tau_h}$ and therefore was not used in the cut-based analysis. Nevertheless, we use this in our ANN analysis. All the relevant feature variables and their definitions are listed below in Table.~\ref{featurevar}.

\begin{table}[htpb!]
\centering

\begin{tabular}{||c ||} 
 \hline
 Variable and Definition \\ [0.5ex] 
 \hline\hline
 $p^{\tau_h}_{T}$ or Transverse momentum of the $\tau$-tagged jet \\ 
 $\slashed{E_T}$ or Missing transverse momenta\\
 $m_T^{\tau_h}$ or Transverse mass \\
 $p_T$ of the leading light-jet \\
 $\Delta\phi(\tau_h,E_T)$ ie. Azimuthal angle between $\tau_h$ and $\slashed{E_T}$ \\ 
 $\Delta\phi(jet,E_T)$ ie. Azimuthal angle between the leading light-jet and $\slashed{E_T}$ \\ 
 $R_{\tau_h}$ or $p_T$ imbalance between $\tau$ decay products\\
 $R_{bb}^{min}$ defined in Equation~\ref{rbbmin}\\
 $p_\theta$  defined in Equation~\ref{ptheta} \\
 Number of light-jets  \\
 Number of $\tau$-tagged jets \\[1ex] 
 \hline
 \end{tabular}

 \caption{\it Feature variables used for training the network.}
  \label{featurevar}
\end{table}

We have used a network with four hidden layers with activation curve relu in each of them. The batch-size is taken to be 1000 and number of epochs per batch is 100. We used 80\% of the dataset for training and 20\% for validation. It is important to avoid over-training of the data sample. Over-training simply means the training sample yields good accuracy but the test sample fails to achieve the similar accuracy. To avoid overtraining we introduced 20\% dropout at each hidden layer.

\begin{figure}[!hptb]
 	\centering
 	\includegraphics[width=9cm,height=7.0cm]{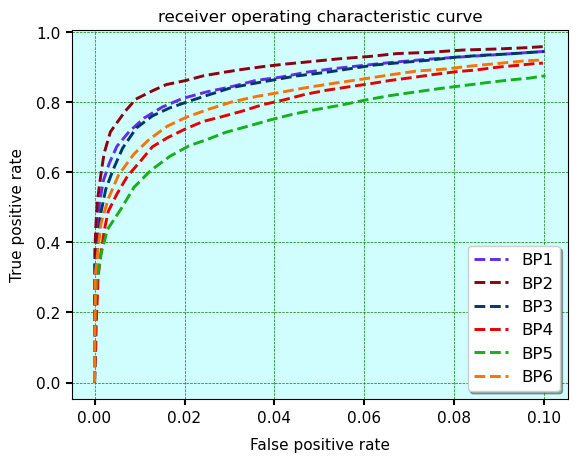}
 \caption{\it ROC curves for various benchmarks.}
 \label{roc}
 \end{figure}

In order to guide the network towards our signal region and to thereby ensure better training, we have applied some weak kinematical cuts, eg. $p_T^{\tau_h} > 50$ GeV and $\slashed{E_T} > 15$ GeV, $m_T^{\tau_h} > 30$ GeV, $R_{\tau_h} > 0.06$  on signal and background events in addition to the pre-selection criteria discussed above. We obtain 92.1\%(BP1), 93.4\%(BP2), 92.2\%(BP3), 90.6\%(BP4), 88.2\%(BP5) and 90.8\%(BP6) accuracy after ANN analysis. In Fig.~\ref{roc}, the Receiver Operating Characteristic (ROC) curve for all the benchmarks are shown.

The area under curve is 0.981(BP1), 0.986(BP2), 0.98(BP3), 0.972(BP4), 0.964(BP5) and 0.974(BP6). We scan over all points on the ROC curve and choose suitable points on the curve, corresponding to each BP that yields the best signal significance. We present the signal significance ${\cal S}$ for all BP's in Table.~\ref{significance_ann}.

\begin{table}[!hptb]
\begin{center}
\begin{tabular}{| c | c | }
\hline
BP1  &  0.5$\sigma$  \\ 
\hline
BP2  &  0.2$\sigma$   \\
\hline 
BP3  &  5.8$\sigma$   \\
\hline
BP4  &  5.2$\sigma$   \\
\hline
BP5  &  26.8$\sigma$   \\
\hline
BP6  &  8.9$\sigma$   \\ 
\hline
\end{tabular}

\caption{\it Signal significance for the benchmark points at 14 TeV LHC with 3000$fb^{-1}$ with ANN. }
\label{significance_ann}
\end{center}
\end{table}

Comparing the results of ANN in Table.~\ref{significance_ann} with that of the cut-based analysis in Table.~\ref{significance} we can see that our analysis with ANN shows significant improvement for all benchmarks. However, the two BP's in Scenario 1, namely BP1 and BP2 perform extremely poorly in cut-based as well as ANN, their signal rate being extremely small compared to the other BP's. BP3, BP4, BP5 and BP6 on the other hand shows remarkable improvement compared to the cut-based results.

\section{Summary and Conclusion}
\label{sec7}

Dark matter has been one of the biggest puzzles of nature since a very long time. Its interactions with known SM fields still remain a mystery. On one hand, much talked about Higgs-portal scenario is getting constrained by data, on the other hand, the idea of an extended scalar sector acting as a portal to the dark sector is becoming more and more popular. The two Higgs doublet models are strong candidates from this viewpoint. Among various two-Higgs-doublet models, the leptophilic 2HDM or Type X 2HDM is quite interesting. It allows for low mass (pseudo)scalars unlike the other types of 2HDM's. It also has enhanced coupling to the leptons especially at large $\tan \beta$. Owing to these particular properties, Type X 2HDM provides solution to the long standing discrepancy between the SM theoretical prediction and experimentally observed value of $g_{\mu}-2$. It is therefore, quite motivating to study the Type X 2HDM with low mass (pseudo)scalars in detail and explore its dark matter as well as collider phenomenology. It has been observed that the dark matter phenomenology of this model is quite different from the other types of 2HDM's, again due to its leptophilic nature. While DM coupling to one doublet gets severely restricted from the requirement of direct detection bounds, coupling to the other doublet remains free to satisfy the observed relic density of the universe. However, a pertinent question then arises, whether the regions of Type X 2HDM, that satisfy the observed muon anomaly or satisfy all the constraints from DM sector and thereby provides a legitimate portal scenario, can be probed at the collider experiments. 

To answer this question, we choose suitable benchmark points from the regions that are of interest from DM phenomenology as well as $g_{\mu}-2$ observation. We select production process and specific final state involving $\tau$-tagged jet + $\slashed{E_T}$ in order to ensure its detectability at the collider. Having chosen relevant kinematic observables we have achieved fair amount of signal-background separation in certain regions of the parameter space. We have also explored $\tau$-polarization observables to reduce polarization sensitive electroweak background. We have found that Scenario 1, where 125 GeV Higgs is the lightest scalar, especially in the wrong-sign region becomes extremely difficult to probe in this channel even at HL-LHC. However, the situation is slightly better in the right-sign region. However, the non-decoupling scenario, where 125 GeV Higgs is the heavier scalar, has the prospect of detection at the HL-LHC. The results improve multifold when we go beyond the cut-based approach and employ the machine-learning techniques such as ANN.
We would like to mention here that in~\cite{Dey:2021pyn} it was found that in the context of Type X 2HDM, the non-decoupling scenario is favored from the viewpoint of high scale validity too. Thus we see that, there exists an intriguing scenario, which can explain the observed muon anomaly, provides a DM candidate and a viable portal mechanism, and is also valid upto high scales and is actually testable at the HL-LHC.

\section{Acknowledgement}

This  work  was  supported  by  funding  available  from  the  Department of Atomic Energy, Government of India, for the Regional Centre for Accelerator-based Particle Physics (RECAPP), Harish-Chandra Research Institute.

\bibliographystyle{jhep}
 \bibliography{ref1.bib}

\end{document}